\documentclass[review,number,sort&compress]{elsarticle}
\usepackage{lineno}
\linenumbers

\usepackage{amssymb}
\usepackage{amsmath}
\usepackage{graphicx}
\usepackage{epstopdf}
\usepackage[figuresright]{rotating}
\usepackage{multirow}

\newcommand{\Ztt}{Z \rightarrow \tau \tau}
\newcommand{\Zee}{Z \rightarrow e e}
\newcommand{\Zmumu}{Z \rightarrow \mu \mu}
\newcommand{\Tthnu}{\tau \rightarrow \tau_h \nu_{\tau}}

\journal{NIM}

\newcommand{\ifm}[1]{\relax\ifmmode #1\else $#1$5\fi}
 
\newcommand{\beq}{\begin{equation}}
\newcommand{\eeq}{\end{equation}}
\newcommand{\beqn}{\begin{eqnarray}}
\newcommand{\eeqn}{\end{eqnarray}}
\newcommand{\bi}{\begin{itemize}}
\newcommand{\ei}{\end{itemize}}
\newcommand{\bd}{\begin{description}}
\newcommand{\ed}{\end{description}}
\newcommand{\bHuge}{\begin{Huge}}
\newcommand{\bhuge}{\begin{huge}}
\newcommand{\bLARGE}{\begin{LARGE}}
\newcommand{\bLarge}{\begin{Large}}
\newcommand{\blarge}{\begin{large}}
\newcommand{\eHuge}{\end{Huge}}
\newcommand{\ehuge}{\end{huge}}
\newcommand{\eLARGE}{\end{LARGE}}
\newcommand{\eLarge}{\end{Large}}
\newcommand{\elarge}{\end{large}}

\def \gtsim    {\relax\ifmmode{\mathrel{\mathpalette\oversim >}}
                  \else{$\mathrel{\mathpalette\oversim >}$}\fi}
\def \ltsim    {\relax\ifmmode{\mathrel{\mathpalette\oversim <}}
                  \else{$\mathrel{\mathpalette\oversim <}$}\fi}
\def\oversim#1#2{\lower4pt\vbox{\baselineskip0pt \lineskip1.5pt
            \ialign{$\mathsurround=0pt#1\hfil##\hfil$\crcr#2\crcr\sim\crcr}}}
\def \dk       {\relax\ifmmode{\rightarrow}\else{$\rightarrow$}4\fi}
\def \to       {\relax\ifmmode{\rightarrow}\else{$\rightarrow$}4\fi}
\def \Dk    {\relax\ifmmode{\Rightarrow}\else{$\Rightarrow$}\fi}
\newcounter{minutes}

\newcommand{\met}{\mbox{${E\!\!\!\!/_{\rm T}}$}}

\def \sp       {\relax\ifmmode{\;}\else{$\;$}\fi}	
\def \PRL      {Phys. Rev. Lett.~}
\def \PR       {Phys. Rev.}
\def \PRD      {Phys. Rev. D}

\def \PLB      {Phys. Lett. B}
\def \ZPC      {Z. Phys. C}	
\def \NPB      {Nucl. Phys. B}
\def \PR       {Phys. Rep.~}

\def \NIM      {Nucl. Instrum. Methods}
\def \NIMA     {Nucl. Instrum. Methods Phys. Res. Sect. A}
\def \RPP	{Rep. Prog. Phys.}
\def \PRP	{Prog. Theor. Phys}
\def \MPL	{{Mod. Phys. Lett.} A}
\def \EPJC	{{Eur. Phys. J.} C}
\def \CPC	{Comput. Phys. Commun.}
\begin{document}

\begin{frontmatter}

\title{Probabilistic Particle Flow Algorithm for High Occupancy Environment}

\author[tamu]{A. Elagin}
\author[fnal]{P. Murat}
\author[lbnl]{A. Pranko}
\author[tamu]{A. Safonov}
\address[tamu]{Texas A\&M University, College Station, TX 77843}
\address[fnal]{Fermi National Accelerator Laboratory, Batavia, IL 60506}
\address[lbnl]{Ernest Orlando Lawrence Berkeley National Laboratory, Berkeley,
CA 94720}

\begin{abstract}
Algorithms based on the particle flow approach are becoming increasingly utilized in collider experiments due to their superior jet energy and missing energy resolution compared to the traditional calorimeter-based measurements. 
Such methods have been shown to work well in environments with low occupancy of particles per unit of calorimeter granularity. However, at higher instantaneous luminosity or in detectors with coarse calorimeter segmentation, the overlaps of calorimeter energy deposits from charged and neutral particles significantly complicate particle energy reconstruction, reducing the overall energy resolution of the method. We present a technique designed to resolve overlapping energy depositions of spatially close particles using a statistically consistent probabilistic procedure. The technique is nearly free of ad-hoc corrections, improves energy resolution, and provides new important handles that can improve the sensitivity of physics analyses: the uncertainty of the jet energy on an event-by-event basis and the estimate of the probability of a given particle hypothesis for a given detector response. When applied to the reconstruction of hadronic jets produced in the decays of tau leptons using the CDF-II detector at Fermilab, the method has demonstrated reliable and robust performance.
\end{abstract}

\end{frontmatter}

\section{Introduction to the Particle Flow Algorithm}

Accurate measurement of the energy of hadronic jets is critical for precision verification of the Standard Model (SM) as well as searches for new physics at current and future collider experiments. A standard jet energy  measurement technique relies on clustering spatially close energy depositions in the calorimeter, the detector designed to measure the energy of particles that produce electromagnetic or hadronic showers in the absorber material. Given that on average about 70\% of a typical jet energy is carried by particles interacting hadronically\footnote{the remaining 30\% is mainly due to neutral pions decaying to pairs of photons, which produce electromagnetic showers.} (mostly $\pi^\pm$, but also $K^{\pm}$, $K^0_L$, protons, neutrons), the resolution of the jet energy measurement is driven by the accuracy  of the hadronic shower energy reconstruction. While the energy of electromagnetic showers can be measured very well, large fluctuations in the development of hadronic showers lead to a significantly lower precision\footnote{A typical example is the CDF calorimeter, which has good electromagnetic calorimeter resolution ${\delta E}/{E} \sim {0.135}/{\sqrt{E}}$ while the response to stable hadrons, e.g. charged pions, is substantially less precise ${\delta E}/{E} \sim {0.5}/{\sqrt{E}}$.}. 
The non-equal response of the non-compensating calorimeters to electromagnetic and hadronic showers\footnote{E.g., main calorimeter systems at ATLAS, CDF, and CMS are all non-compensating.} further biases the overall jet energy scale and degrades the resolution. Special corrections accounting for non-equal response can only partially recover this reduction in resolution. While the presence of many particles in a jet averages out fluctuations in the measurement of energy of individual hadronic showers, jet energy resolution remains poor for jets of low ($\sim$10--30 GeV) and moderate ($\sim$30--60 GeV) energies. Incidentally, the resolution of low-to-moderate energy jets has a strong impact on the sensitivity of many physics analyses performed at hadron colliders, from precision measurements of the Standard Model parameters to searches for the Higgs boson in $b\bar{b}$ and $\tau \tau$ channels and searches for new phenomena such as predicted by Supersymmetry. Mismeasurements of the jet energy also bias the measurement of the missing transverse energy ($\met$) in an event, a key discriminant used in many analyses searching for new phenomena, calculated as an imbalance of the energy in the event in the direction transverse to the beam line. Enhancing the discovery potential of current and future collider experiments is therefore motivating the development of improved jet energy measurement techniques.

A significant improvement in the jet energy resolution at hadron collider experiments has been achieved with the deployment of a technique known as the Particle Flow Algorithm (PFA), e.g. see ~\cite{cms_tau}. PFA achieves better jet energy resolution by reconstructing and measuring energies of individual particles in a jet using information from several detector sub-systems. For example,the  momenta of charged hadrons can be measured much more accurately using the tracking system (except for the case of very high transverse momenta, which is not relevant for this discussion), than in the calorimeter. This allows one to replace the less accurate calorimeter measurement of the energy carried by charged hadrons in the PFA jet energy calculation with: 
\begin{eqnarray}
E_{jet} = \displaystyle\sum\limits_{tracks } E_{trk} +  \displaystyle\sum\limits_{\gamma's} E_{\gamma} + \displaystyle\sum\limits_{n} E_{n}, 
\end{eqnarray}
where the first term is the energy of the charged particles in the jet, the second term accounts for energy of photons accurately measured in the electromagnetic calorimeter, and $E_n$ is the energy of stable neutral hadrons, e.g. neutrons or $K^0_L$'s, which still relies on the hadron calorimeter. The corresponding relative jet energy resolution can be written in terms of single particle relative resolutions as:
\begin{eqnarray}
\label{jetenergy_error}\frac{\sigma^2(E_{jet})}{E_{jet}^2} =  \frac{1}{E_{jet}^2} \times \left( { \displaystyle\sum_{tracks} E_{trk}^2} \frac{\sigma^2(E_{trk})}{E_{trk}^2} + { \displaystyle\sum_{\gamma's} E_{\gamma}^2} \frac{\sigma^2(E_{\gamma})}{E_{\gamma}^2} + { \displaystyle\sum_{n's} E_{n}^2} \frac{\sigma^2(E_{n})}{E_{n}^2} \right) 
\end{eqnarray}
Note that only the last term depends on the potentially poor calorimeter resolution for the energy of hadronic showers. However, because the fraction of the jet energy carried by stable neutral hadrons is on average only around 10\%, its contribution to the overall jet energy uncertainty is strongly suppressed by $\sum E_n/E_{jet}$. With the remaining 90\% of energy accurately measured either in the tracker or in the electromagnetic calorimeter, the PFA-based jet energy reconstruction can substantially outperform the traditional calorimeter-only based measurements. Furthermore, the bias in the energy scale related to calorimeter non-compensation effects is significantly reduced as it is only present in the suppressed third term.

Apart from an obvious pre-requisite of highly efficient tracking, the performance of a PFA-based reconstruction in a realistic setting depends critically on one's ability to correctly identify and separate calorimeter energy depositions from spatially close particles. One example illustrating the issue is an overlap of energy deposits in the calorimeter due to a charged pion and a neutron. In this case one has to ``guess'' the fraction of the measured calorimeter energy deposited by the charged pion, so that the excess can be attributed to a neutral hadron. The dependence of the jet energy resolution on the overlap effects is sometimes parameterized by amending Eq.~(\ref{jetenergy_error}) with the so called ``confusion term''~\cite{confusion_term}  $\sigma^2_{conf}$. The relative importance of the confusion term depends on the power of the algorithm and the detector design features, but it generally increases with the coarser calorimeter segmentation and higher particle densities. In extreme cases, the large size of the confusion term can completely eliminate the advantages of the PFA over traditional calorimeter-based measurements.

PF-based algorithms were successfully implemented at LEP in the 1990's ~\cite{aleph_pfa} and have been pursued in developing the physics program at the International Linear Collider (ILC)~\cite{ilc_pfa}. At hadron collider experiments, a simplified version of a PFA-based algorithm was implemented for the reconstruction of hadronically decaying tau leptons at CDF at the end of Run I~\cite{amy_thesis}.
In hadronic decays, a tau lepton decays into a neutrino and one or more charged and neutral hadrons\footnote{the number of charged hadrons in tau lepton decays is always odd owing to the conservation of electric charge}. While tau decays often proceed via intermediate resonances, e.g. $\rho$ or $a_1$, the final stable charged hadrons are usually pions and, less often, kaons. Neutral hadrons produced in hadronic tau lepton decays are dominated by neutral pions\footnote{Only about 2\% of the time tau decay products contain one or more neutral kaon}, which promptly decay to photons via $\pi^0 \to \gamma \gamma$.  Tau leptons accessible at hadron machines typically originate from $Z$ or $W$ decays and therefore have a substantial momentum. This leads to their decay products being fairly collimated and appearing as jets of charged and neutral hadrons, reminiscent of regular jets originating from quarks or gluons. The decay products of tau leptons undergoing hadronic decays are therefore frequently referred to as ``tau jets''.
The PF-based reconstruction allowed for a strong improvement in the energy resolution of hadronic tau jets, and the technique was further improved and used at CDF for Run II analyses~\cite{cdf_tau}. A more comprehensive implementation of the same technique~\cite{cdf_pfa} has been shown to improve the generic jet resolution at CDF compared to a calorimeter only reconstruction. 
However, the coarse segmentation of the CDF calorimeter has led to a sizable confusion term associated with the substantial probability for more than one particle to deposit energy in a given calorimeter cluster. The difficulty in resolving such overlaps of energy depositions required to reconstruct momenta of individual particles allowed for only a limited improvement. A complete PFA algorithm developed by the CMS experiment~\cite{cms_tau} has allowed for a strong improvement in the jet energy measurement as well as a more accurate missing transverse energy scale and resolution. The CMS detector is well suited for PFA-based reconstruction due to the fine granularity of the electromagnetic calorimeter and the longitudinal profiling of hadronic showers in the central part of the detector, which improves their spatial resolution. However,  the series of the ``High Luminosity LHC'' upgrades are expected to result in significant increases in particle occupancies per event. Maintaining high performance of the PFA-based reconstruction in the new regime requires the development of techniques capable of efficiently resolving energy overlaps.


In this paper, we discuss the challenges and implications of deploying a PFA-based reconstruction in an environment with frequent energy overlaps (Section 2). In Section 3 we present a technique designed to resolve the overlapping energy depositions of spatially close particles using a statistically consistent probabilistic procedure. In addition to improving the energy resolution, the technique allows for combining measurements from multiple detectors, as opposed to ``substituting'' one measurement with another in existing algorithms. It is nearly free of ad-hoc corrections, thus minimizing distortions due to the discontinuities of the correction functions. The algorithm has additional unique features, such as the ability to calculate the jet energy uncertainty on a jet-by-jet basis, and provides the measure of the overall consistency of the measurement, improving the sensitivity of physics analyses. In Section 4, we describe the implementation of this technique for reconstructing the energy of tau jets, the decay products of hadronically decaying tau leptons, at CDF and illustrate its performance in a realistic setting using the actual experimental data in Section 5. 
\section{Challenges of the High Occupancy Environment}

The reconstruction of hadronically decaying tau jets at CDF is a good example of a problem with frequent overlaps of energy deposits from nearby particles. The CDF calorimeter has projective tower geometry with azimuthal segmentation $\phi = 15^{\circ}$ and pseudorapidity segmentation $\eta \approx 0.1$ and provides very limited information about the lateral and longitudinal shower profiles\footnote{As discussed further in the text, there is a strip-wire chamber embedded inside the electromagnetic calorimeter at $\sim6X_{0}$, which allows for rough measurements of the latteral profile in some cases. Longitudinal profile information is limited to two energy measurements for deposits in the electromagnetic and hadron compartments of a tower.}. With a typical angular size of a hadronic tau jet being of the order of 0.05-0.1 rad, there is a substantial probability for several or even all particles within the tau jet to cross the face of the calorimeter within the boundaries of a single calorimeter tower. Treatment of frequent energy overlaps is therefore a key consideration in designing a PFA-based reconstruction at CDF. 

To set the stage, we need to briefly describe the sub-detector systems used in tau reconstruction and identification, a full description of the CDF-II detector is available elsewhere~\cite{CDFdet}. The CDF tracking system provides nearly 100\% efficient tracking within the pseudorapidity range of $|\eta| < 1$, which is relevant for tau reconstruction. Its main element is the Central Outer Tracker (COT), a drift chamber that covers radii from 0.4 m to 1.37 m immersed in a 1.4 T magnetic field, providing momentum resolution of $\delta p_{T}/p_{T}^2 \approx 0.0017 ($GeV$/c)^{-1}$. If available, hits from the silicon vertex detector (SVX) are added to the COT information, further improving the resolution. Central electromagnetic (CEM) and hadronic (CHA) calorimeters cover the pseudorapidity region of $|\eta| < 1.1$. CEM is a lead-scintillator calorimeter with resolution $\delta E_{T}/E_{T} = 0.135/ \sqrt{E_{T}} \oplus 0.02$. CHA is an iron-scintillator calorimeter with the single pion energy resolution of $0.5/ \sqrt{E_{T}} \oplus 0.03$. Both calorimeters have a projective tower geometry with tower size $\Delta \phi \times \Delta \eta \approx 15^{\circ} \times 0.1$ and neither of the calorimeters measures either the longitudinal or lateral shower profile. The Shower Maximum (CES) detector, consisting of a set of strip-wire chambers embedded inside the CEM at the expected maximum of the electromagnetic shower profile, enables measurement of the position of electromagnetic showers with an accuracy of a few mm by reconstructing clusters formed by strip and wires. While rarely used to measure energy of the electromagnetic showers, CES cluster's pulse height provides a measurement of electromagnetic shower energy with the resolution of $\delta E/E = 0.23$ for showers due to energetic photons or electrons. As pulse heights of the one-dimensional strip and wire clusters reconstructed for the same shower are typically within $\approx 7$\% of each other\footnote{The CES energy resolution is driven by the fluctuations in the amount of ionization produced inside the CES chambers and not by the measurement of the charge collected on strips and wires}, multiple showers within a single CES chamber can typically be correctly reconstructed by matching the 1D strip and wire clusters using their pulse heights. The much broader hadronic showers frequently extend over multiple CHA towers and their spatial position can only be inferred from the energy measured in each tower. Early hadronic showers can deposit part of their energy in CEM and produce signals in CES, which sometimes complicates the reconstruction of CES clusters, e.g. if overlapping with showers produced by photons (from $\pi^0\rightarrow\gamma\gamma$) in the same jet.

\begin{figure}[htb]
\vspace{8pt}
\begin{center}
\includegraphics[width=0.48\linewidth]{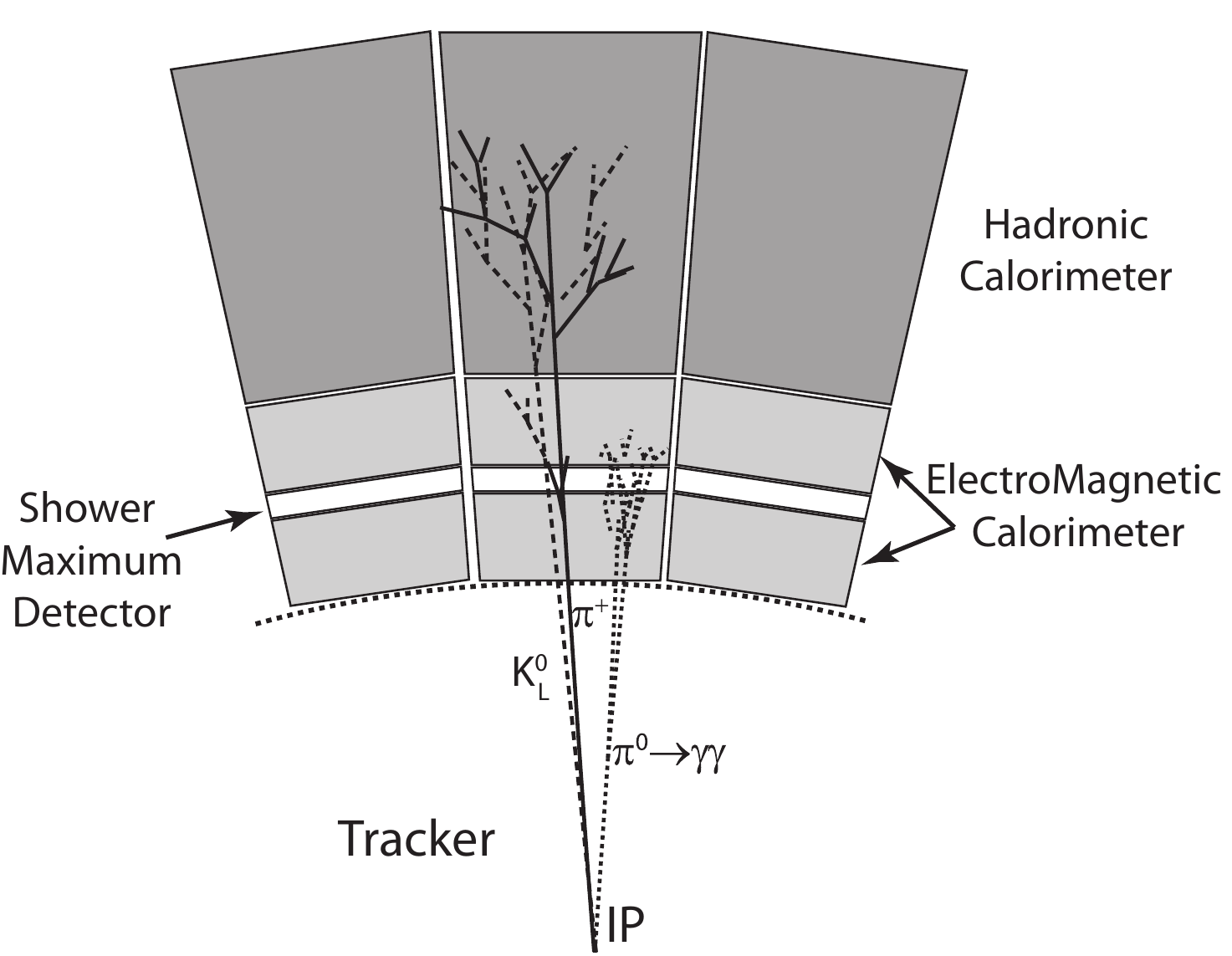}
\end{center}
\caption{Illustration of the measurements available in the CDF-II detector setup for a hypothetical narrow jet of particles consisting of a charged pion (trajectory and the shower shown as solid lines), a neutral pion (dotted line) decaying to two unresolved photons and a neutral hadron (dashed line).}
\label{event-diagram}
\end{figure}

\begin{figure}[htb]
\includegraphics[width=0.48\linewidth]{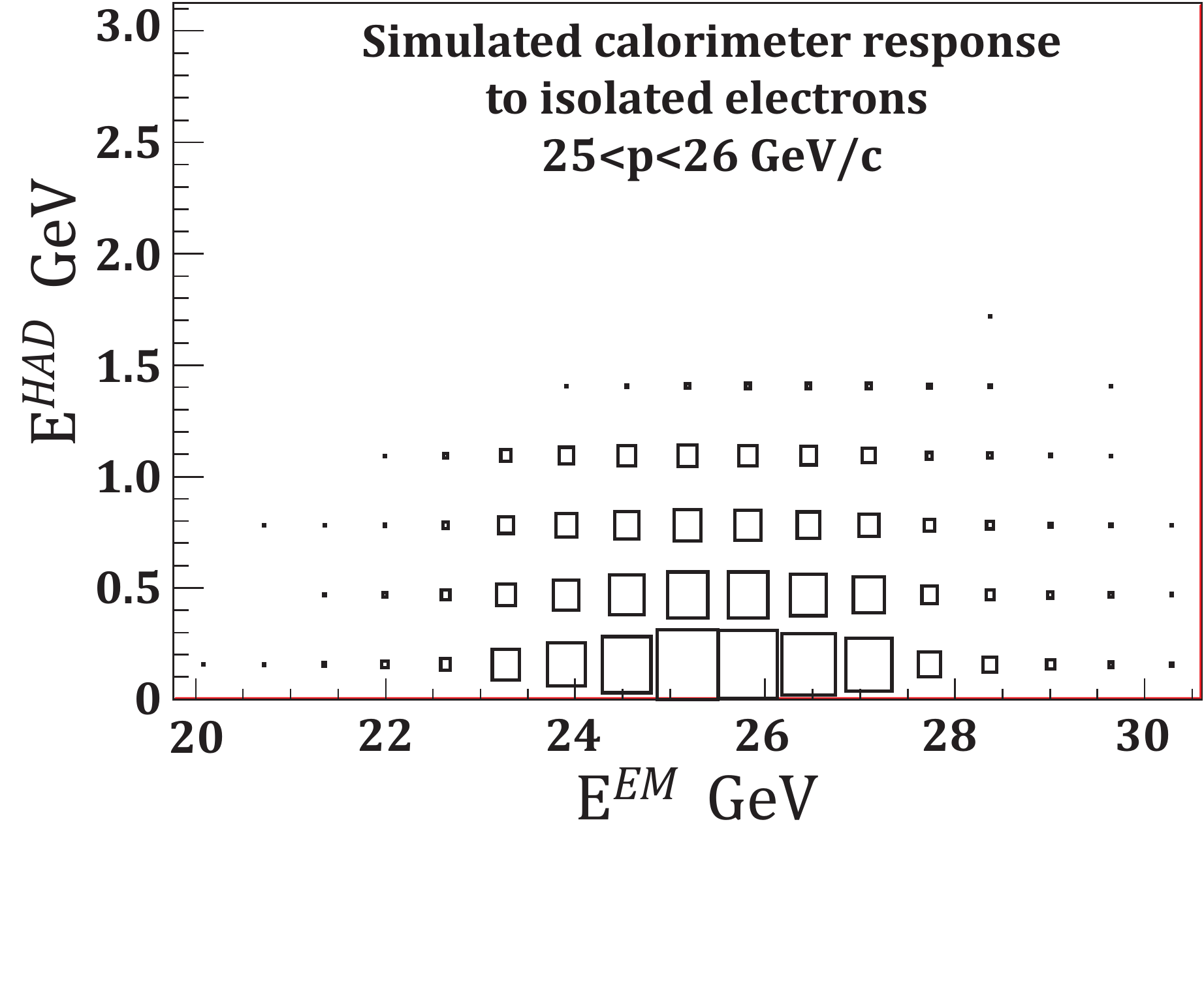}
\includegraphics[width=0.48\linewidth]{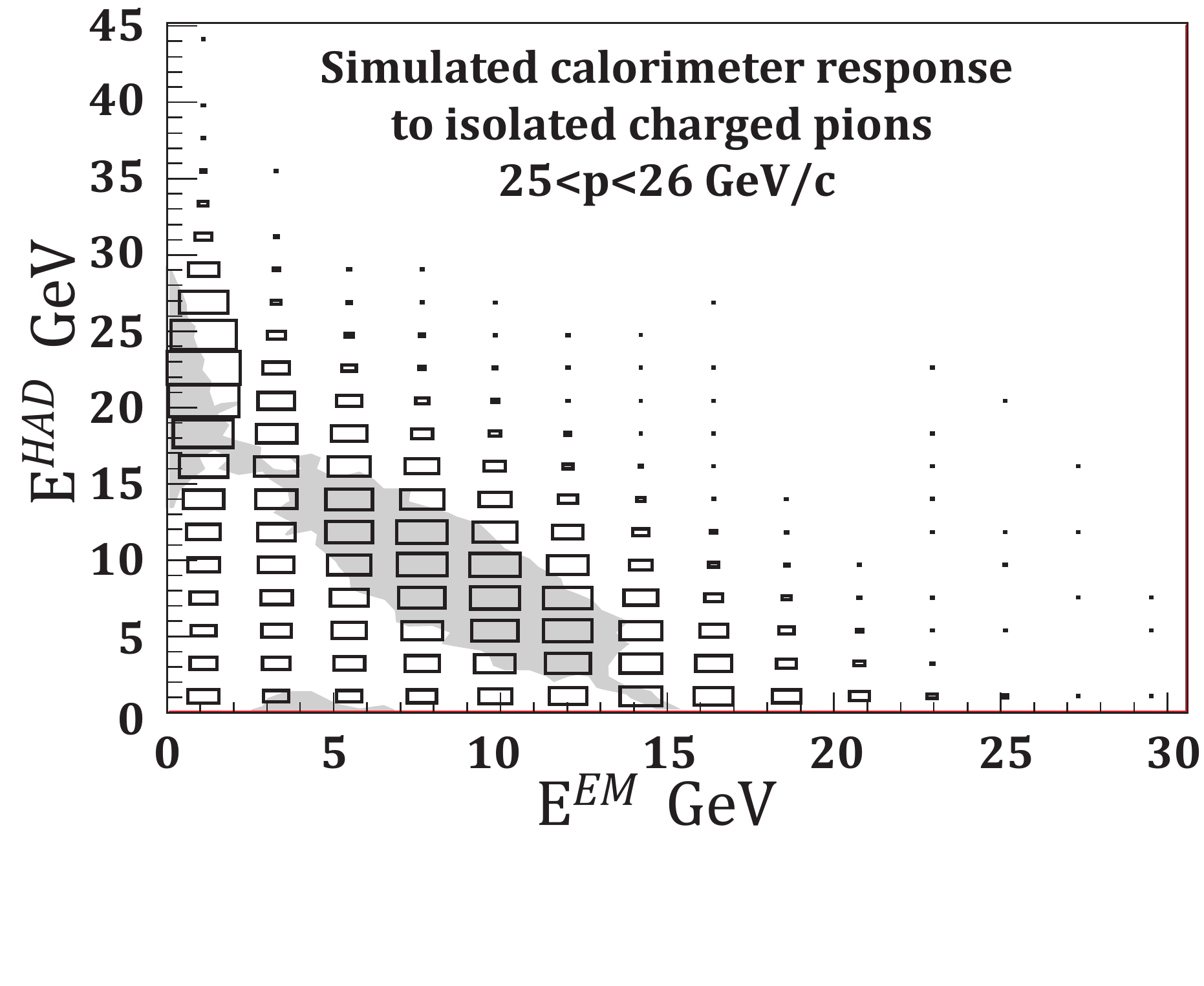}
\caption{Examples of the calorimeter response for (a) simulated isolated electrons with $p=25$ GeV/c and (b) simulated isolated charged pions with $p=25$ GeV/c in the plane $E^{EM}$ versus $E^{HAD}$. The size of the boxes in the plots is proportional to the probability density; the shaded area indicates the area of the highest density as obtained from the same distribution plotted with finer bin size.}
\label{RespFunc1}
\end{figure}

Let us consider a relatively simple example of a jet containing a charged pion $\pi^+$, a neutral pion $\pi^0$ decaying to two unresolved photons $\gamma_1 \gamma_2$ (depositing energy in a single tower), and possibly a neutral hadron $n$, as illustrated in Fig.~\ref{event-diagram}. While the $\pi^+$ momentum is known from the tracker, the energy estimation for neutral particles relies on the calorimeter measurement. However, the energy registered in the electromagnetic and hadronic parts of the calorimeter, $E^{EM}_{meas}$ and $E^{HAD}_{meas}$, is a sum of the unknown deposits by each of the particles in the jet, including that by the charged pion:
\begin{eqnarray}
\label{example_eqn_em} E^{EM}_{meas} = E^{EM}_{\pi^+} + E^{EM}_{\gamma_1 \gamma_2} + E^{EM}_{n} \\
\label{example_eqn_had} E^{HAD}_{meas} = E^{HAD}_{\pi^+} + (E^{HAD}_{\gamma_1 \gamma_2})+ E^{HAD}_{n},
\end{eqnarray}
resulting in an under-constrained system with two equations and six unknowns. As the leakage of the electromagnetic showers from photons into the hadron calorimeter is typically small, as illustrated in Fig.~\ref{RespFunc1}(a) showing $E^{EM}$ vs. $E^{HAD}$ for simulated electrons, the corresponding term $E^{HAD}_{\gamma_1 \gamma_2}$, shown in parentheses in Eq.(\ref{example_eqn_had}), can be neglected. While it reduces the number of unknowns, solving the system of Eqs.~(\ref{example_eqn_em},\ref{example_eqn_had}) requires disentangling contributions from hadronically interacting particles. While $E^{EM}_{\pi^+}$ and $E^{HAD}_{\pi^+}$ terms are correlated with the accurately measured momentum of $\pi^+$, the correlation is not trivial, as illustrated in Fig.~\ref{RespFunc1}(b) showing the 2D distribution of $E^{EM}$ vs. $E^{HAD}$ for a simulated sample of charged pions with $p_{\pi^+}=25$ GeV/c. The complex shape of the dependence owes to the large fluctuations in the development of hadronic showers and the non-compensating nature of the CDF calorimeter. As $E^{EM}_{\pi^+}$ cannot be reliably estimated, and $E^{EM}_{n}$ is completely unconstrained, the momentum of the $\pi^0$ cannot be calculated directly. Estimating the jet energy directly in the PFA approach is therefore hampered by two issues: (i) difficulty in estimating $E^{EM}$ for hadronically interacting particles, required to evaluate the $\pi^0$ momentum, and (ii) difficulty in estimating $E^{HAD}_{\pi^+}$, required to estimate the momentum of $n$. Measuring the momentum of a combined $\pi^0+n$ system, e.g. by ``guessing'' the charged pion energy depositions and assigning the rest to the $\pi^0+n$ system, is nearly exactly equivalent to measuring the jet energy using the calorimeter only thus negating all advantages of the PFA technique.

An algorithm based on solving Eqs.(\ref{example_eqn_em},\ref{example_eqn_had}) directly with the specific purpose of reconstructing hadronic tau jets was implemented in the ``tracks+$\pi^0$'s'' algorithm at CDF and used in the early Run-II analyses. The idea was to simplify the problem by assuming the absence of neutral hadrons and estimate $E^{EM}_{\pi^+}$ as an average energy deposition in the EM calorimeter for a charged pion with given momentum (measured in the tracker). The remaining portion of the measured electromagnetic energy can then be taken as the energy of the $\pi^0$ (Eq.(\ref{example_eqn_em})). Alternatively, one can assume that charged pions always behave in the electromagnetic calorimeter as minimal ionizing particles. While delivering a significant improvement over the calorimeter-only measurement for a large fraction of events, the algorithm featured long tails in the energy resolution. These tails have been traced to jets with several particles depositing energy in the same calorimeter tower. In physics analyses, an underestimation of the energy of quark or gluon jets containing neutral hadronically interacting particles also leads to an increase in background contamination. Additional corrections based on detecting incompatibilities of the reconstructed energy with the initially unused Eq.(\ref{example_eqn_had}) or gross disagreements with the low resolution measurement of $\pi^0$ energy in the Shower Maximum detector allow for a reduction of the tails in the energy resolution. However, the ad-hoc nature and complexity of the corrections, as well as the algorithm's inability to consistently treat correlations and incorporate other available measurements motivate developing a more comprehensive method. 

\section{PPFA: The Probabilistic Particle Flow Algorithm}

The challenge of solving an underconstrained system with significant correlations and additional redundant measurements outlined in previous section can be addressed with a probabilistic approach. For every hypothesis of the jet particle content  (the number of particles of each type), one can define a probability estimator (likelihood) for a set of particles of given type and momenta to result in a particular set of detector measurements. These measurements could represent energy counts in calorimeter towers, cluster energies, track momenta or any other available measurement. The likelihood can be written as follows: 
\begin{eqnarray}
\label{pffa_likelihood_general} {\cal L}(\vec{p}\; | \vec{E}_{meas}) = \int{{\cal M}(E_{1}^{1}, ... , E_{i_p}^{j_m},  E^1_{meas}, ..., E^{j_m}_{meas}) \times \displaystyle\prod_{i\; j} {\cal P}_{ij}(E_i^j | p_i) dE_i^j},
\end{eqnarray}
where index $i$ runs over particles in a jet ($i = 1, ..., i_p$), $p_i$ is the true momentum of particle $i$, $E^j_{meas}$ stands for each available measurement ($j=1, ..., j_m$ runs over all available measurements), ${\cal P}_{ij}(E_i^j | p_i)$ is the ``response function'' for particle $i$ with true momentum $p_i$, to produce a contribution $E_i^j$ to a measurement $j$ (if the particle cannot contribute to the particular measurement $j^\prime$, we use ${\cal P}_{i{j^\prime}}(E_i^{j^\prime} | p_i)=\delta(E_i^{j^\prime})$), and ${\cal M}$ contains information about correlations between contributions of each particle to each measurement. One example of the latter is the correlation between the deposits of energy $E_i^j$ in an electromagnetic calorimeter tower $j$ by all particles crossing it, in which case ${\cal M}$ will contain a product of expressions of the form $\delta(\displaystyle\sum_{i}{E_{i}^j}-E^j_{meas})$ for each relevant tower $j$. The summation runs over all particles $i$ that cross this tower and the delta function ensures that the integration is performed over the parameter space where the assumed contributions to the measured energy by individual particles sum up into the experimentally measured energy deposition in that tower. Another example is the correlation between the energy deposited by particle $i$ in the electromagnetic calorimeter tower $j_1$ and a hadron calorimeter tower $j_2$ it crosses. In this case ${\cal M}$ would have to account for the correlation between the energy deposits $E_i^{j_1}$ and $E_i^{j_2}$ in the two towers. Once such a global likelihood function is constructed, $\vec{p}^0$ corresponding to its maximum will determine the most probable set of particle momenta, thus achieving the goal of fully reconstructing the event using all available detector information. The type of each particle and their number can be taken as parameters of the global likelihood, allowing one to also determine the most probable particle content of a jet.

While building a global and fully inclusive likelihood is certainly possible, it is hardly practical. However, this approach can be deployed to solve specific problems like measuring jet energies in environments with frequent energy overlaps in the calorimeter. Here, we will describe an example of one such possible PPFA implementation. For simplicity, this example will use the energy of pre-reconstructed  electromagnetic and hadronic calorimeter clusters as the basic measurements $E^j_{meas}$, but an implementation using tower energy measurements would be very similar. The PPFA probability for a set of particles with momenta $p_i$ to produce a set of calorimeter measurements $E^j_{meas}$ for each cluster $j$ in electromagnetic or hadron calorimeter can be written as follows:
\begin{eqnarray}
\label{pffa_likelihood} {\cal L}_p(\vec{p} \; | \vec{E}_{meas}) = \int{ {\cal M^\prime} \displaystyle\prod_{j} \delta(\displaystyle\sum_{i} E_i^j -E^j_{meas}) \times \displaystyle\prod_{i\; j}{\cal P}_{ij}(E_i^j | p_i) dE_i^j},
\end{eqnarray}
where $\vec{p}$ is the vector of particle momenta $p_i$, index $i$ runs over the list of particles in a jet, index $j$ runs over the available measurements (in our example, the electromagnetic and hadronic calorimeter's cluster energy measurements), $E^j_i$ is the contribution to the measured energy in cluster $j$ by $i^{th}$ particle, $E^j_{meas}$ is the measured energy for cluster $j$, and the response function ${\cal P}_{ij}(E_i^j | p_i)$ is the probability for particle $i$ with true momentum $p_i$ to deposit energy $E_i^j$ in cluster $j$ (${\cal P}_{ij}$ depends on the type of particle), and ${\cal M^\prime}$ describes correlations that remain unaccounted after the delta functions have been introduced. The likelihood ${\cal L}_p$ is essentially a sum of probabilities of all possible outcomes, i.e. the specific values of energy deposited by each particle in the electromagnetic and hadronic calorimeter clusters, consistent with the actual cluster energy measurements (the latter is ensured by the delta functions). The probability of each outcome is a product of probabilities  ${\cal P}_{ij}$ for each particle to deposit given amounts of energy $E_i^{j_1}$, $E_i^{j_2}$ in the hadronic and electromagnetic calorimeters, given their assumed true momenta $p_i$. In many practical cases, the strongest effect that ${\cal M^\prime}$ in Eq.(\ref{pffa_likelihood}) has to properly account for is the correlation of the values of the deposits by the same particle in the electromagnetic and hadronic calorimeters, e.g. early showering of a charged hadron can lead to a larger than typical deposition of energy in the electromagnetic calorimeter, but the energy deposited in the hadron calorimeter would consequently be lower than typical. The easiest way to take this kind of correlation into account is to switch to two-dimensional response functions ${\cal P}^{CAL}(E^{EM_{j_1}}_i, E^{HAD_{j_2}}_i | p_i)$, where $j_1$ and $j_2$ are the indices of the electromagnetic and hadronic calorimeter clusters the particle traverses. In this case, ${\cal M^\prime}$ is no longer needed and the distributions shown in Figs.~\ref{RespFunc1}(a) and (b) can be normalized and used as response functions ${\cal P}^{CAL}(E^{EM}, E^{HAD} | p)$ for electrons and charged pions, respectively. For numerical calculations in practical applications, one should first simplify the expression for ${\cal L}_p$ by integrating over some of the variables to remove delta functions. In some cases it is as easy as dropping the integration over $dE^j_i$ with the combinations of indices $i$ and $j$ corresponding to cases where particle $i$ is making no contribution to measurement $j$, which is equivalent to integrating over the corresponding $dE^j_i$ and taking into account that ${\cal P}_{ij} (E^j_i | p_i)=\delta({E^j_i})$ in the above equation. In other cases, one needs to choose which variables to integrate over to optimize the speed and accuracy of the numeric calculations.

\begin{figure}[htb]
\vspace{8pt}
\includegraphics[width=0.48\linewidth]{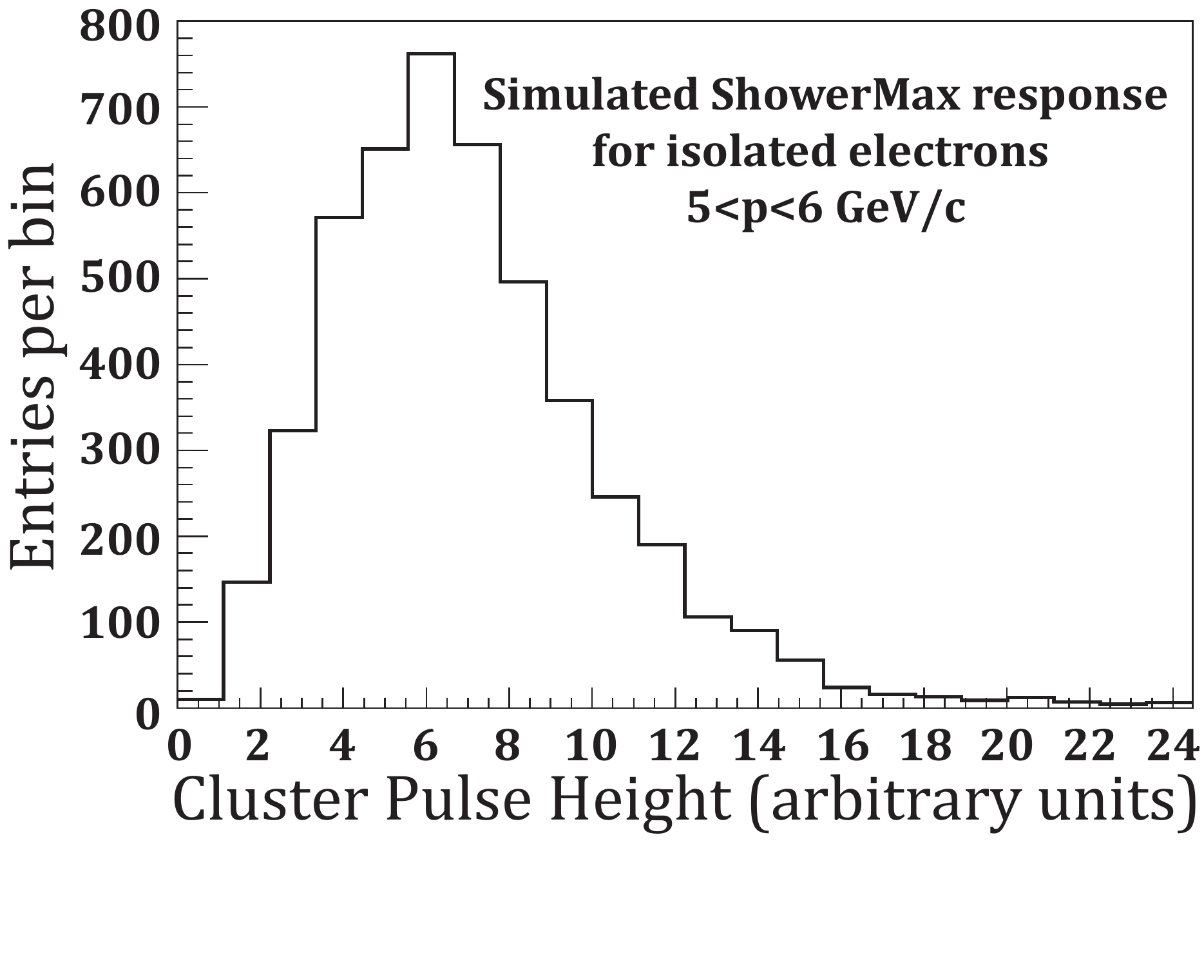}
\includegraphics[width=0.48\linewidth]{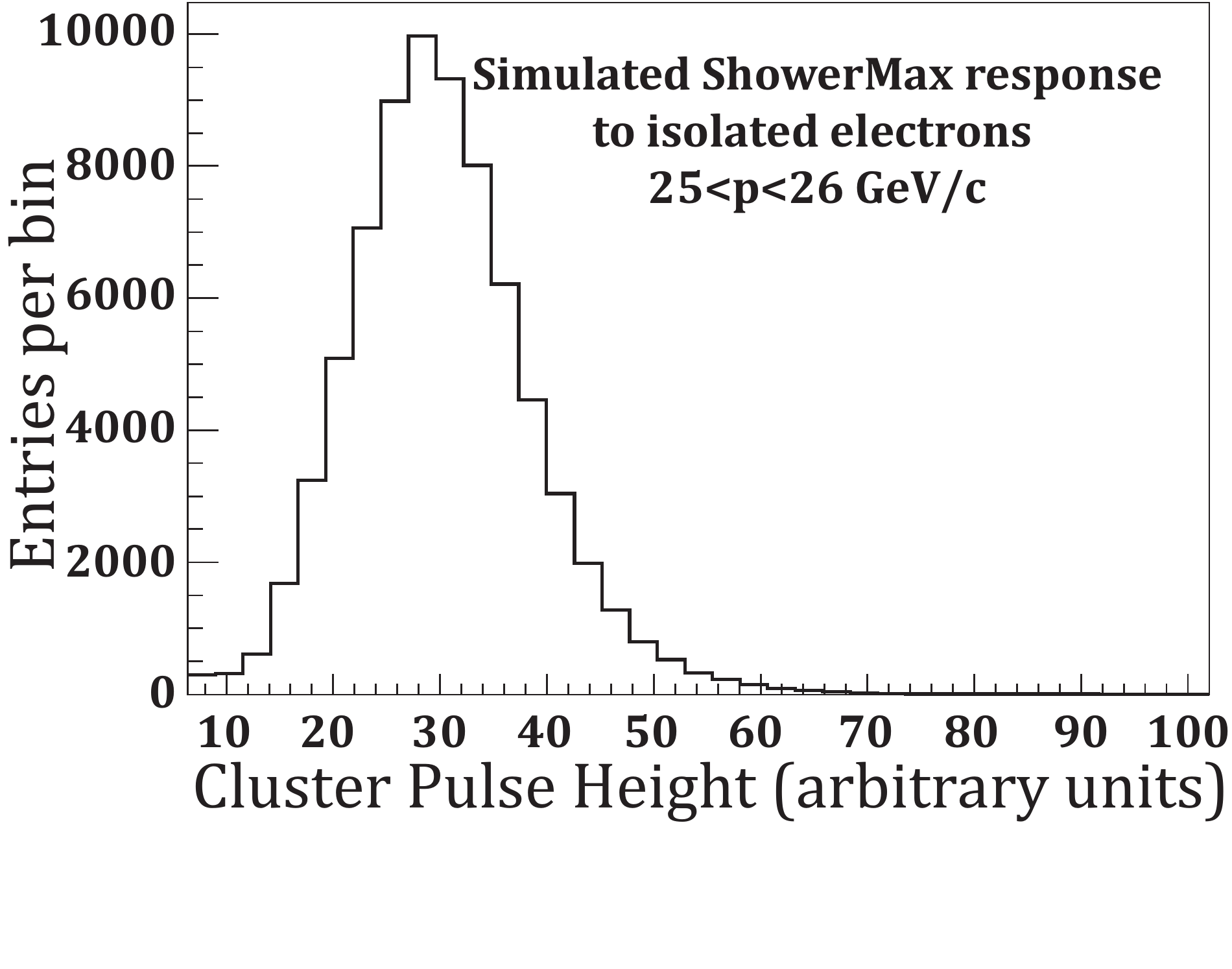}
\caption{Examples of the Shower Maximum detector response functions for simulated isolated electrons with momenta in the ranges $p=5-6$ GeV/c (left) and $p=25-26$ GeV/c (right). Similarity in the response between electrons and photons allows using these functions in constructing likelihood functions for either electrons or photons.}
\label{RespFunc}
\end{figure}

Additional measurements can be easily incorporated by modifying the likelihood function with Bayesian-like ``priors''. For example, information from tracking or Shower Maximum detectors can be added by multiplying the initial likelihood function by a probability to measure a certain track momentum or pulse height given the assumed true momentum of charged pions, electrons or photons. For example, distribution shown in Fig.~\ref{RespFunc} upon normalization can be used as the response functions of the Shower Maximum detector ${\cal P}^{CES}_{\gamma}(E^{CES} | p_{\gamma})$ for photons with momenta ranges  $p=5-6$ and $25-26$ GeV/c. The inclusion of ``priors'' is equivalent to expanding the list of measurements in the original likelihood function and introducing further correlation information into the matrix ${\cal M}$.

While the most probable set of particle momenta $\vec{p^0}$ is obtained by maximizing the likelihood ${\cal L}_p(\vec{p} \; | \vec{E}_{meas})$, the likelihood shape in the $\vec{p}$ space can be used to evaluate the uncertainty in the energy determination for each particle. If one primarily seeks to measure the energy of a particular jet in the event as often is the case, one can use the likelihood defined in Eq.(\ref{pffa_likelihood}) to obtain a ``posterior'' distribution for the jet energy, defined as a sum of the energies of the particles assigned to the jet. This is accomplished by integrating over $d\vec{p}=dp_1 ... dp_{i_p}$
\begin{eqnarray}
\label{pffa_likelihood_e} {\cal L}_E(E_{jet} | \vec{E}_{meas}) =\int{\cal L}_p(\vec{p} \; | \vec{E}_{meas}) \times \delta(\displaystyle\sum_{i=1}^{N} p_{i}- E_{jet}) d\vec{p}, 
\end{eqnarray}
and we have assumed here that the first $N$ particles in the list are those assigned to the jet in question. In the presence of correlations,  ${\cal L}_E$ may provide a more convenient estimate of the jet energy and its uncertainty. The shape of the jet energy ``posterior'' allows for the estimation of the uncertainty in the measured jet energy. If one needs to simultaneously evaluate the energy of several jets in the event, Eq.(~\ref{pffa_likelihood_e}) needs to be modified by introducing additional delta-functions (one per jet) with the summation running over indices of particles assigned to each of the jets.

Once the most likely set of particle momenta $\vec{p^0}$ is found, one can further test the ``goodness'' of the particle hypothesis. We define a p-value as the probability to observe a combination of detector measurements $\vec{E}^\prime_{meas}$ that is equally or less likely than the actual set $\vec{E}_{meas}$ observed in the event, given that the true combination of particles and momenta is the one that maximizes the likelihood in Eq.~(\ref{pffa_likelihood}):
\begin{eqnarray}
\label{p-value} p(\vec{p}^0, \vec{E}_{meas}) = \frac{\displaystyle\int_{ {\cal L}_p(\vec{p}^0 \; | \vec{E}^\prime_{meas}) \leq {\cal L}_p(\vec{p}^0 \; | \vec{E}_{meas})}{  {\cal L}_p(\vec{p}^0 \; | \vec{E}^\prime_{meas}) d\vec{E}^\prime_{meas} }}
{\displaystyle\int{  {\cal L}_p(\vec{p}^0 \; | \vec{E}^\prime_{meas}) d\vec{E}^\prime_{meas} }}
\end{eqnarray}
The p-value can be easily calculated numerically by generating ``pseudo-experi-ments,'' in which one generates ``pseudo-deposits'' of energy by each particle with momenta $p^0_i$ towards each cluster energy measurement using the same response functions. The sum of the deposits of all particles crossing particular clusters yields a set of pseudo-measurements $\vec{E}^{\prime}_{meas}$. The probability of the generated outcome is given by ${\cal L}_p$, and the integrated probability of observing equally or less probable set of measurements than the actually observed $\vec{E}_{meas}$ gives the p-value. A too low p-value may indicate that the initial particle hypothesis should be modified. Note that interpreting measured p-values has to be done carefully as the arbitrary addition of new particles to make the observed calorimeter response ``perfect'' may degrade the resolution by biasing the measurement towards the calorimeter-based jet energy measurement. While the p-value defined in Eq.(~\ref{p-value}) is global for the entire event, a p-value can be defined for each individual jet or a set of particles. For example, one can either build ${\cal L}_p$ by only include particles of interest, e.g. the ones belonging to a particular jet, or by integrating the global ${\cal L}_p$ over a subset of momenta $p_i$ belonging to particles that are of no interest for a given measurement.
\section{PPFA-Based Reconstruction of Hadronically Decaying Tau Leptons at CDF}

In this section we describe a practical implementation of the method developed for hadronic tau jet reconstruction at CDF. In the following, we discuss the CDF baseline hadronic tau jet reconstruction, which is used as a starting point for the algorithm. We then discuss the PPFA strategy, measurement of the response functions, mathematical definition of the PPFA likelihood function and the ``p-value,'' and the algorithm used for correcting the initial particle hypothesis. We conclude with evaluating the algorithm's energy resolution using simulation. As we are primarily interested in improving energy reconstruction for tau jets, we implement a ``local'' version of the PPFA, with the definitions presented in the previous section only including particle candidates that contribute to a particular jet and completely ignoring the rest of the information contained in the event. 

\subsection{Baseline Hadronic Tau Jet Reconstruction at CDF}
The construction of hadronic tau jet candidates at CDF starts with selecting continuous clusters of calorimeter towers. The clustering starts with a ``seed tower,'' defined as any tower with $E_T>5$ GeV/c and at least one track with $p_T>5$ GeV/c pointing to the cluster. Broad clusters with more than six contiguous calorimeter towers with $E_T>1$ GeV/c are excluded from consideration as true tau jets  almost always result in narrow clusters of just a few towers. Clusters outside the central part of the detector ($|\eta| \leq 1$) are also discarded to ensure a high tracking efficiency for the remaining candidates. Given the size of the CDF calorimeter towers of $\Delta \eta \times \Delta \phi \sim 0.1 \times 0.25$, the efficiency of the calorimeter-related selections is very high, reaching nearly 100\% for hadronically decaying taus with visible $p_T>10$ GeV/c. The seed track $p_T$ requirement brings a non-negligible inefficiency for tau jets of low-to-moderate visible momentum, but its strong power in rejecting quark and gluon jet backgrounds made it a standard in all CDF analyses involving hadronic tau jets. Next, all tracks within a signal cone of $\Delta R = \sqrt{\Delta \phi^2 + \Delta \eta^2} < 0.17$ around the seed track are associated with the tau candidate.

\subsection{Implementation Strategy}
The likelihood-based PPFA algorithm starts with the initial hypothesis that every reconstructed track is a charged pion, every reconstructed cluster in the Shower Maximum detector with no track pointing to it is a photon, and no other particles are present in the jet. While this initial hypothesis can be corrected at a later point in the algorithm, in most cases it turns out to be true owing to the low rate of the track and Shower Maximum reconstruction failures and the low branching fraction of hadronic tau lepton decays for modes with neutral hadrons except $\pi^0$'s, e.g. $\tau \to K_L +X $. Next, we define the probability function using pre-calculated response functions (details for both are discussed in the following two sub-sections) and perform a scan in the multi-dimensional parameter space of momenta of the particles, assumed to comprise the hadronic tau jet, searching for the maximum of the likelihood function. 

After the most likely combination of particle momenta is determined, we construct the ``p-value'' which measures the probability that the given particle content and momenta hypothesis result in detector measurements less or equally as likely as the observed response. If the p-value is too low, the particle content hypothesis is modified by adding a photon, which is assumed to be not reconstructed either due to the detector inefficiency or an overlap with a track (Shower Maximum cluster will be vetoed if it is reconstructed too close to the extrapolated position of a charged track), and the full calculation is repeated. If the p-value remains too low, the particle content is modified by adding a stable neutral hadron ($K_L$) and the likelihood calculation is repeated. The procedure continues until an acceptable outcome is achieved or after running out of the pre-set options.

\subsection{Response Functions of the CDF Detector Sub-systems}
As discussed earlier, the relevant detector measurements include tracking, measurements of energy deposited in the electromagnetic and hadronic calorimeter towers, and the measured CES cluster energy. Because the precision of the CDF tracking is much higher than the accuracy of other measurements, the tracker response function for charged pions as a function of pion momenta can be safely approximated by a delta function to simplify further calculations. To determine the calorimeter response functions for charged pions, we use the CDF GEANT-3~\cite{geant3} based simulation package tuned using the test beam data. Isolated charged pions are selected using hadronic tau decays $\tau^\pm \to \pi^\pm \nu_\tau$ from an inclusive $Z/\gamma^*\to\tau\tau$ simulated sample of events generated with Pythia~\cite{pythia}. We calculate response functions for charged pions with momenta ranging from 1 to 100 GeV/c in steps of 1 GeV/c. Large fluctuations in the development of hadronic showers and their large lateral size, frequently spanning across several CHA towers, make it impractical to calculate responses separately for each tower in a multi-tower cluster. Instead, we measure the hadronic calorimeter response for charged pions by summing tower energies in a square of $3\times3$ towers centered on the extrapolated position of the $\pi^\pm$ track. In the CEM, hadronic showers rarely deposit energy in more than a single tower, therefore the charged pion electromagnetic deposition is calculated using the energy in the tower pointed at by the track associated with $\pi^\pm$. To take into account the strong correlation of the energy depositions by the same particle in CEM and CHA, we define a 2-dimensional response function in the $E^{EM}$ versus $E^{HAD}$ plane. Fig.~\ref{RespFunc1}(b) shows an example of the calorimeter responses in CEM and CHA for simulated isolated charged pions with momenta $25< p_{\pi}< 26$ GeV/c. The adequacy of the CDF simulation of the calorimeter response can be inferred from the results of a dedicated study~\cite{cdf-calo-nim}, in which simulation predictions were compared with the pion test beam data and with the collisions data using a pure sample of isolated charged pions. When normalized to unity, these response functions represent the probability density functions (PDF) for a charged pion with a particular momentum to produce a given response in the calorimeter, which we will refer to as ${\cal P}^{CAL}_\pi (E^{EM}, E^{HAD} | p_\pi)$.

The vast majority of photons in tau jets originate from $\pi^0 \to \gamma \gamma$ and typically have energy of the order of a few GeV, making accurate understanding of the calorimeter response for low energy photons particularly important. While the response functions for photons can be measured directly from the simulation, validating them with the data can be difficult owing to the challenges in selecting a high purity sample of low energy photons in data. Fortunately, the calorimeter response to photons and electrons is nearly identical, allowing for the use of a relatively high purity sample of electrons in data obtained by tagging photon conversions. Similar to the case of charged pions, we calculate 2-dimensional response functions for photons with momenta ranging from 1 to 100 GeV/c in steps of 1 GeV/c in the $E^{EM}$ versus $E^{HAD}$ plane. Fig.~\ref{RespFunc1}(a) shows an example of the calorimeter response function for photons with the true momenta $25< p_{\gamma}< 26$ GeV/c. We denote the response functions of this type as ${\cal P}^{CAL}_\gamma (E^{EM}, E^{HAD} | p_\gamma)$.

As mentioned earlier, the CES energy measurement is used in the likelihood function as, despite its modest resolution, it can help correctly assign energies in difficult cases. As photon candidates reconstructed in CES have highly correlated strip and wire pulse heights, we only use the strip based measurements to determine the energy of a given CES cluster. Examples of the CES response functions ${\cal P}^{CES}_\gamma (E^{CES}| p_\gamma)$ for isolated photons with energies $5 < p_{\gamma} < 6$ GeV and $25 < p_{\gamma} < 26$ GeV are shown in Figs.~\ref{RespFunc}(a) and (b), respectively.

\subsection{Computation of the PPFA Likelihood}
In our implementation, the initial particle hypothesis assumes each reconstructed track to be due to a charged pion and each reconstructed CES cluster not associated with a track to be due to a photon (or perhaps two merged photons, which makes little difference). The tracking momentum measurement is taken to be exact due to the superior resolution of the CDF tracker. To include calorimeter measurements, the highest $p_T$ track associated to a tau candidate is extrapolated to the CES radius and the corresponding calorimeter tower becomes a seed tower. A grid of 3x3 towers is formed around the seed tower, and each track and CES cluster is associated to one tower on the grid. Each electromagnetic tower provides its own measurement $\vec{E}^{EM}_{meas}$ (components of this vector will be denoted as ${E}^{EM_j}_{meas}$ $j=1,...,9$) used in the likelihood.  For the hadronic calorimeter, we sum the energies of all nine towers into a single measurement, $E^{HAD}_{meas}=\sum{E^{HAD_j}_{meas}}$, for the entire "super-cluster". Under the assumption that the decay products of a tau jet are charged tracks and photons, the likelihood function has the following form:
\begin{eqnarray}
\nonumber {\cal L}_p(\vec{p}_\pi, \vec{p}_\gamma, \vec{p}_n | \vec{E}^{EM}_{meas}, E^{HAD}_{meas}, \vec{E}^{CES}_{meas}) =\\  
{\nonumber \int{\delta(\displaystyle\sum_{i=1}^{N_\gamma} E_{\gamma_i}^{HAD}+\displaystyle\sum_{k=1}^{N_\pi} E_{\pi_k}^{HAD} +\displaystyle\sum_{l=1}^{N_n} E_{n_l}^{HAD} - E^{HAD}_{meas}) \times}} \\
{\nonumber \displaystyle\prod_{j=1}^9 dE_{\gamma_i}^{EM_j}   dE_{\pi_k}^{EM_j} dE_{n_l}^{EM_j}  
\delta(\displaystyle\sum_{i=1}^{N_\gamma} E_{\gamma_i}^{EM_j}+\displaystyle\sum_{k=1}^{N_\pi} E_{\pi_k}^{EM_j} +\displaystyle\sum_{l=1}^{N_n} E_{n_l}^{EM} - E^{EM_j}_{meas})} \times \\
{\nonumber \displaystyle\prod_{i=1}^{N_\gamma} dE_{\gamma_i}^{HAD}  \displaystyle\prod_{k=1}^{N_\pi}   dE_{\pi_k}^{HAD} 
\displaystyle\prod_{l=1}^{N_n}{dE_{n_l}^{HAD}} \; \;  {\cal P}^{CAL}_{\gamma}(E_{\gamma_i}^{EM_j}, E_{\gamma_i}^{HAD} | p_{\gamma_i}) \times} \\
\label{cdf_likelihood} {\cal P}^{CAL}_{\pi}(E_{\pi_k}^{EM_j}, E_{\pi_k}^{HAD} | p_{\pi_k}) \; {\cal P}^{CAL}_{n}(E_{n_l}^{EM_j}, E_{n_l}^{HAD} | p_{n_l}) 
\; {\cal P}^{CES}_{\gamma}(E_{\gamma_i \; meas}^{CES} | p_{\gamma_i}),
\end{eqnarray}
where the integration runs over all possible depositions of energy by each individual particle in each available calorimeter measurement, the delta functions in the second line ensure that the sum of the deposits for each measurement is equal to the observed value, and the third line includes response functions for photons, charged pions and neutral hadrons in the calorimeter and in the CES detector. One can choose to convert Eq.(\ref{cdf_likelihood}) into a posterior probability distribution to estimate the hadronic tau jet energy as:
\begin{eqnarray}
\nonumber {\cal L}_E(E_{jet} | \vec{E}^{EM}_{meas}, E^{HAD}_{meas}, \vec{E}^{CES}_{meas}) =\int{\cal L}_p(\vec{p}_\pi, \vec{p}_\gamma | \vec{E}^{EM}_{meas}, E^{HAD}_{meas}, \vec{E}^{CES}_{meas}) \\
\label{cdf_likelihood_e} \times \delta(\displaystyle\sum_{i=1}^{N_\gamma} p_{\gamma_i}+\displaystyle\sum_{k=1}^{N_\pi} p_{\pi_k} + \displaystyle\sum_{l=1}^{N_n} p_{n_l} - E_{jet}) d\vec{p}_\pi d\vec{p}_\gamma d\vec{p}_n
\end{eqnarray}

\begin{figure}[htb]
\includegraphics[width=0.46\linewidth]{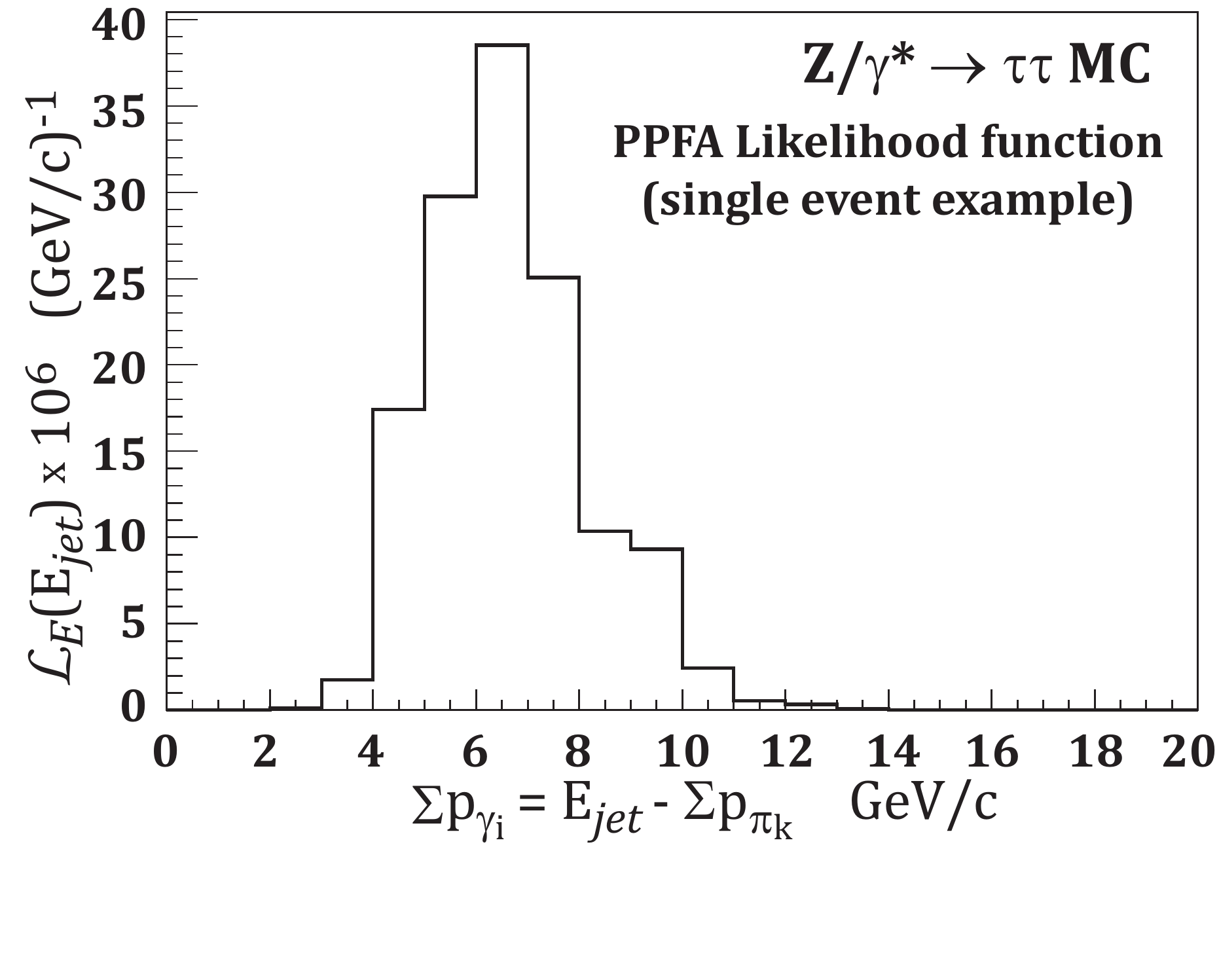}
\includegraphics[width=0.45\linewidth]{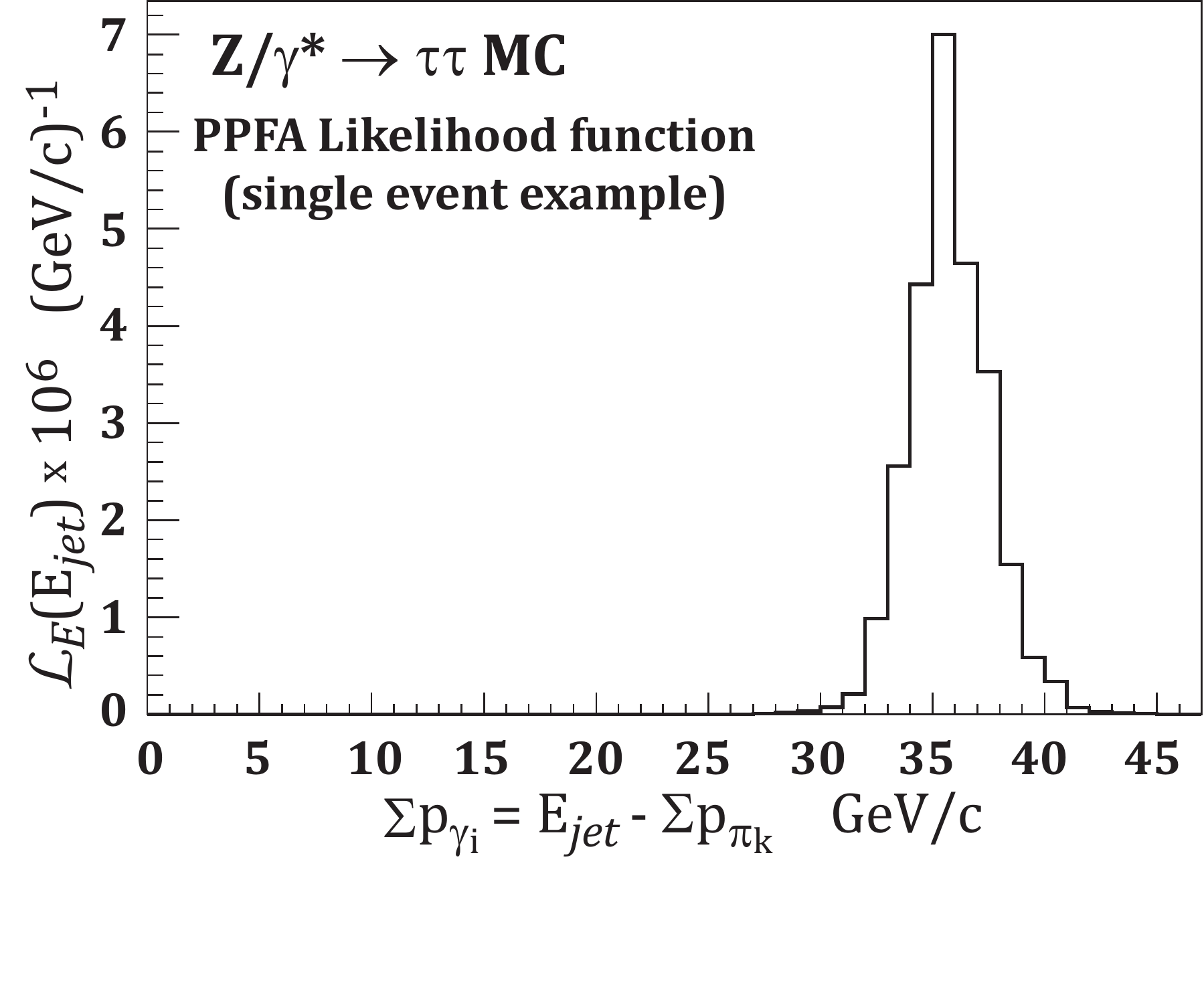}
\caption{Examples of ${\cal L}_E(E_{jet})$ for two representative simulated $\Ztt$ events plotted versus combined energy of the photon candidates in the jet related to $E_{jet}$ via $\sum p_\gamma = E_{jet}-\sum p_\pi$. These distributions serve as statistical probability density functions for the values of the measured energy for a given jet. In the PPFA implementation discussed in this paper, the maximum of the ${\cal L}_E(E_{jet})$ distribution is used as the estimator for the jet energy, while the width and the shape of the distribution yield the uncertainty in the measurement of the jet energy on a jet-by-jet basis.}
\label{fig_likelihoods}
\end{figure}

While the integral form presented in Eqs.(\ref{cdf_likelihood},\ref{cdf_likelihood_e}) appears fairly complicated, it is straightforward to implement in the code and compute numerically using the Monte Carlo integration technique. Values of $\vec{p}_{\pi_k}$ and $\vec{p}_{\gamma_i}$, which maximize ${\cal L}(\vec{p}_\pi, \vec{p}_\gamma)$ in Eq.(\ref{cdf_likelihood}) represent the best estimate for energies of particles produced in the tau decay, under the assumption that the initial hypothesis about the particle content was correct. Figure~\ref{fig_likelihoods} shows examples of the ${\cal L}_E(E_{jet})$ distributions for two representative events from a sample of simulated $Z \to \tau \tau$ events. 

\subsection{The Reduced p-Value Definition}
Photon reconstruction failures or the presence of a stable neutral hadron, e.g. $K^0_L$, may lead to an incorrect initial particle hypothesis. Such occurrences result in a suboptimal estimation of the energy, and therefore it important to detect and correct such cases. We define a p-value using Eq.(\ref{p-value}), but, to speed up the calculations, we do two simplifications to the definition of the likelihood ${\cal L}_p$ in Eq.~(\ref{cdf_likelihood}). First, because in practice most of the cases affected by the incorrect initial hypothesis can be identified through inconsistencies between the available calorimeter and tracker measurements, we drop the terms associated with the CES. Second, we combine the nine electromagnetic towers in the hadronic tau cluster into a single ``super-tower'' with energy $E^{EM}=\sum{E^{EM_j}}$, where the summation runs over the nine towers, and define the ``reduced'' version of Eq.~(\ref{cdf_likelihood}):
\begin{eqnarray}
{\nonumber {\cal L}_p^\prime (\vec{p} \; | (E^{EM}_{meas}, {E}^{HAD}_{meas}) = 
\displaystyle\int{\delta(\displaystyle\sum_{m=1}^{9} E^{EM_m}_{meas} - E^{EM}_{meas})} }  \\
\label{L-p-red} \times {\cal L}_p(\vec{p} \; | \vec{E}^{EM}_{meas}, {E}^{HAD}_{meas}) \displaystyle\prod_{j=1}^{9} dE^{EM_j}_{meas} 
\end{eqnarray}

\begin{figure}[tbh]
\includegraphics[width=0.45\linewidth]{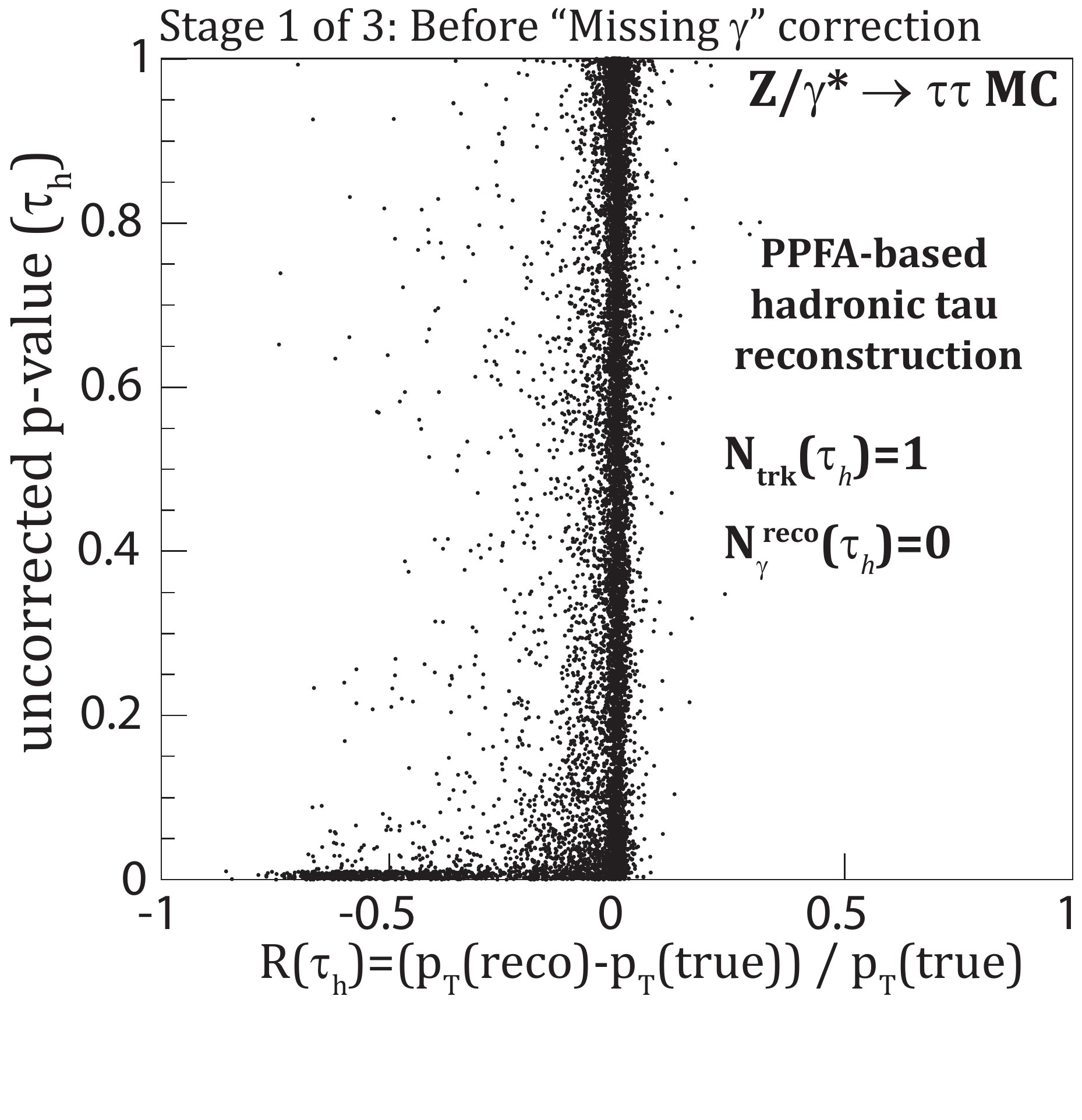}
\includegraphics[width=0.45\linewidth]{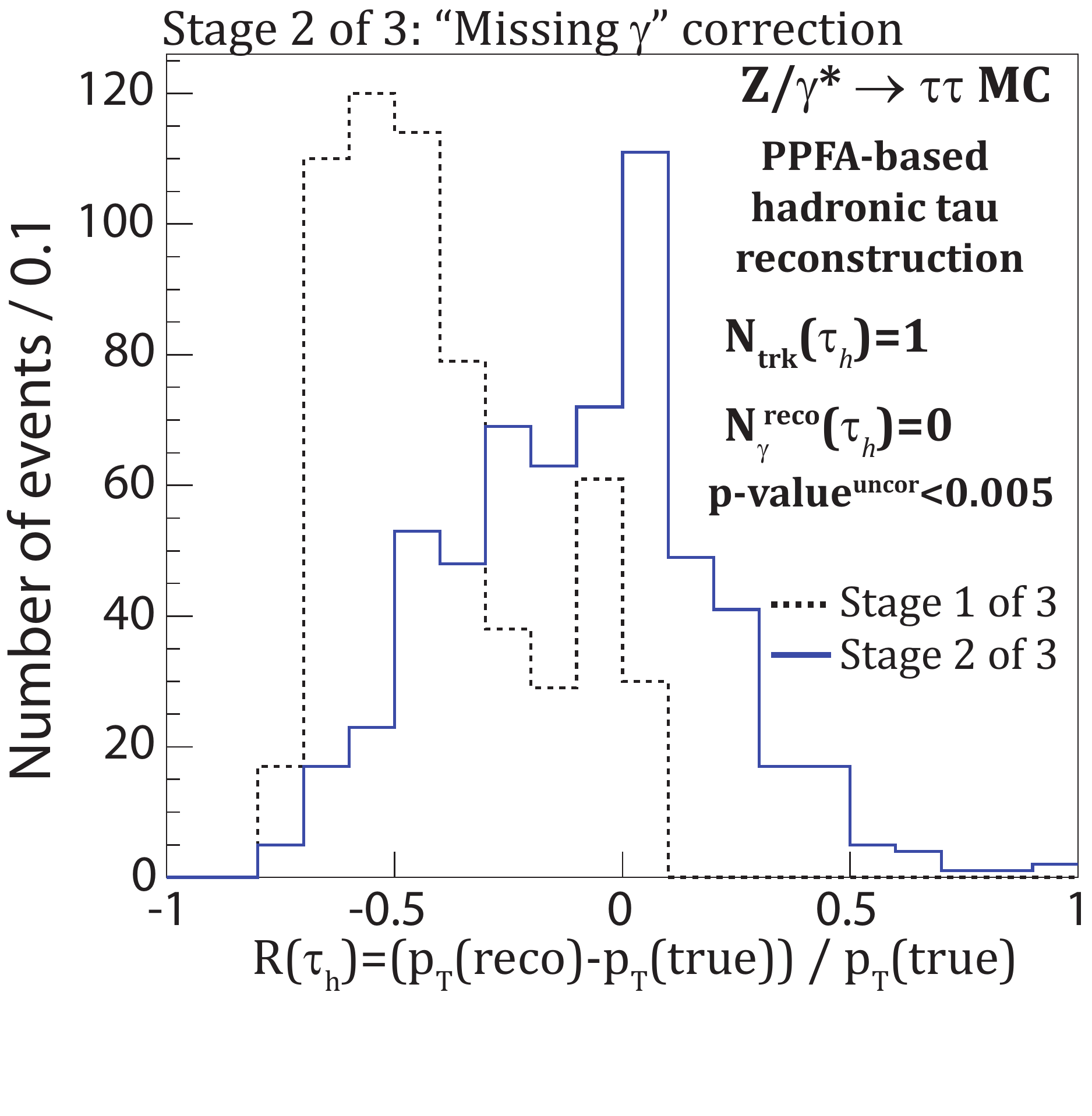}
\caption{$\Ztt$ events in CDF II detector simulation: 1-prong taus with no
photon candidate reconstructed by CES. Left: p-value versus relative energy mismeasurement $R(\tau_h)$.
Right: $R(\tau_h)$ for events with small p-value before correction (dashed black
line) and after correction for missing photon (solid blue line). As points in the enhancements near the x-axis have typically 
very low p-values ($10^{-3}$ or less), coarser y-axis binning is chosen to keep these enhancements visible.}
\label{PhotonCorrection}
\end{figure}

We then define the ``reduced'' p-value according to Eq.~(\ref{p-value}) using the reduced ${\cal L}^\prime_p$. This p-value quantifies how frequently a set of particles with true momenta $\vec{p}^0$ can produce a set of measurements equally or less probable than the one observed in data. The p-value is sensitive to inconsistencies in the available calorimeter measurements and can be used to detect mistakes in the initial particle content hypothesis. Figure~\ref{PhotonCorrection}(a) shows the distribution of the reduced p-value for all reconstructed hadronic tau jets in the sample of simulated $Z \to \tau \tau$ events. The p-value is plotted as a function of the relative difference between the reconstructed visible tau jet energy at the maximum of the likelihood function and the true visible jet energy obtained at the particle generator level. It is evident that a vast majority of mismeasured jets have very low reduced p-value. As it will be shown next, most of these mismeasurements owe to the incorrect initial particle hypothesis.

\subsection{Corrections to the Particle Content Hypothesis}
Based on the simulation studies, the majority of mismeasurements owing to the incorrect initial particle hypothesis fall into two categories. The first category includes tau jets with one charged pion and typically one $\pi^0$, where none of the photons were reconstructed in the CES. This can happen for one of the following three reasons: (i) a simple CES reconstruction failure (either dead channels or a photon mostly properly registering in the EM calorimeter but landing outside the fiducial volume of CES), (ii) the CES cluster is vetoed due to being too close to the extrapolated track position, or (iii) photon(s) falling into the uninstrumented regions (``cracks'') between the calorimeter $\phi$-wedges. The last case is likely to be impossible to correct as the deposited electromagnetic energy is highly sensitive to small differences in the electromagnetic shower development. In addition, photons hitting the cracks may deposit a substantial portion of their energy in the hadron calorimeter. All three cases lead to a substantial underestimation of the tau jet  energy as only the momentum of the track would count towards the measurement. To correct for this effect we apply the following procedure: if a tau candidate with a single reconstructed track and no reconstructed photons has the reduced p-value that is too small ($p<0.005$), we first attempt to correct it by introducing an additional photon. As no CES measurement is available for this photon, the term with ${\cal P}^{CES}$ in Eq.(\ref{cdf_likelihood}) is removed and the likelihood function with modified particle hypothesis ${\cal L}({p}_\pi,{p}_\gamma)$ (or the corresponding ${\cal L}_E$) is recalculated. The new energy is taken as the updated energy of the tau jet. Figure~\ref{PhotonCorrection}(b) shows the relative difference between the reconstructed and the true values of the jet energy for these jets before and after the correction. While the improvement is evident, the catastrophic cases where photons hit the cracks between the calorimeter wedges cannot be fully recovered and contribute to reduced resolution. Another contribution, which makes the distribution broader, comes from events in the second category which are discussed next and can be corrected. 

\begin{figure}[tbh]
\includegraphics[width=0.45\linewidth]{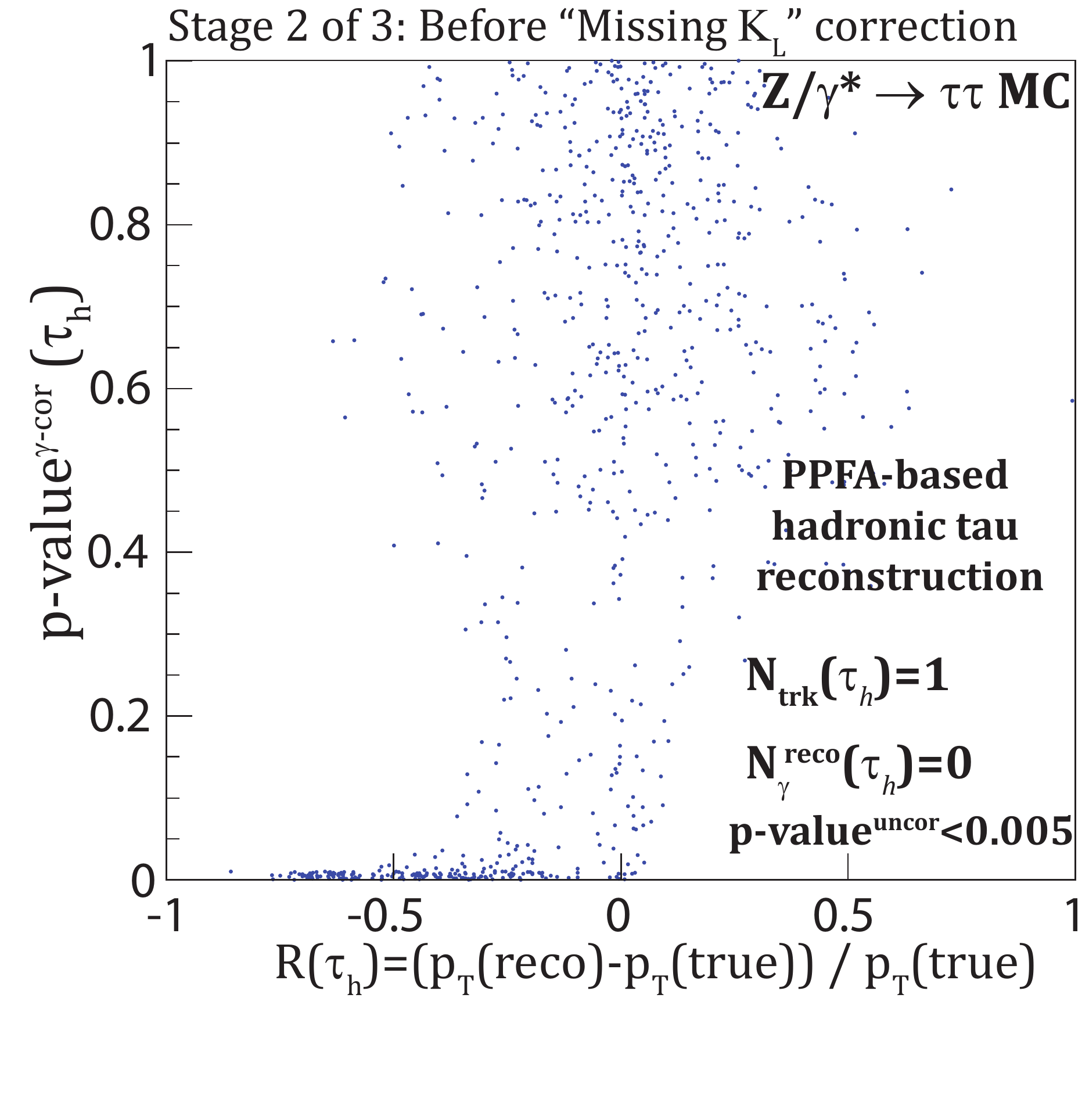}
\includegraphics[width=0.45\linewidth]{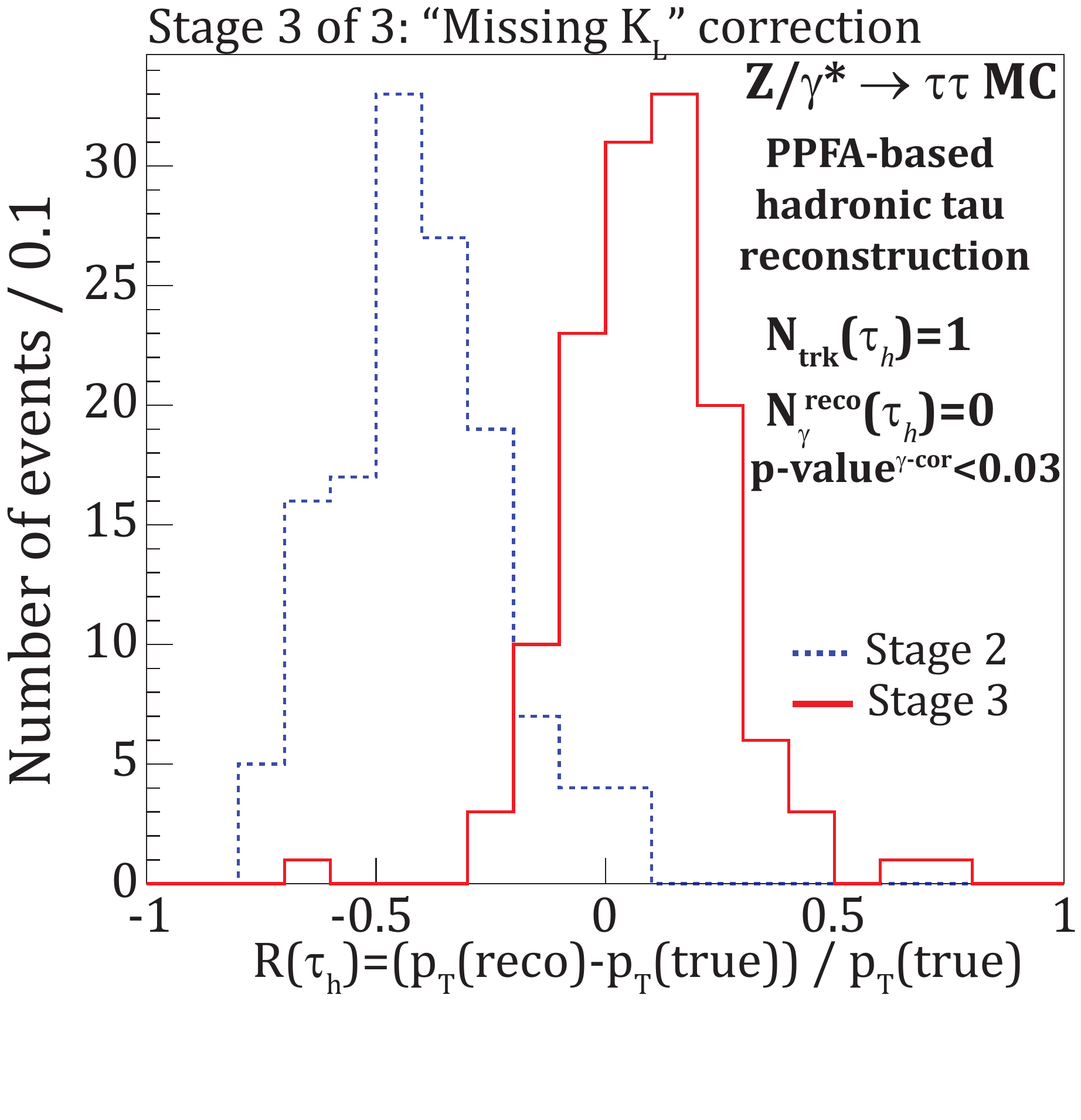}
\caption{$\Ztt$ events in CDF II detector simulation: 1-prong taus with no
photon candidate reconstructed by CES and p-value$^{\rm uncor}<0.005$. Left:
p-value after photon correction versus $R(\tau_h)$.
Right: $R(\tau_h)$ for events with small p-value$^{\rm \gamma-cor}$ before kaon correction (dashed blue
line) and after correction for kaons (solid red line).}
\label{KaonCorrection}
\end{figure}

Tau jets with one charged hadron and a stable neutral hadron (kaon), which is not included in the initial particle content hypothesis, typically have an excess of energy measured in the hadron calorimeter compared to what one would expect from a single charged pion. As the excessive energy in the hadron calorimeter detected using the p-value cannot be accounted for by adding a photon at the previous step, the p-value for these jets remains small after an attempted correction of the initial particle content hypothesis, as shown in Fig.~\ref{KaonCorrection}(a). Therefore, for jets with exactly one reconstructed track and no reconstructed photons that had a low initial p-value ($p<0.005$) and continue to have a low p-value after the photon correction (the threshold is $p<0.03$), the particle content hypothesis is modified to contain one charged pion and one neutral kaon. Technically, it is accomplished by adding a term ${\cal P}_n^{CAL} (E^{EM}, E^{HAD} | p_n) = {\cal P}_\pi^{CAL} (E^{EM}, E^{HAD} | p_n)$ (as the calorimeter response for charged pions and neutral hadrons is very similar) in Eq.(\ref{cdf_likelihood}), and adjusting the argument of the delta-functions to include a new particle. The energy of the tau jet candidate is updated with the energy obtained from maximizing ${\cal L}_p({p}_\pi, p_n)$ (or the corresponding ${\cal L}_E$). The relative difference between the reconstructed and the true tau jet energy before and after the correction for this class of jets is shown in Fig.~\ref{KaonCorrection}(b).

\begin{figure}[tbh]
\includegraphics[width=0.30\linewidth]{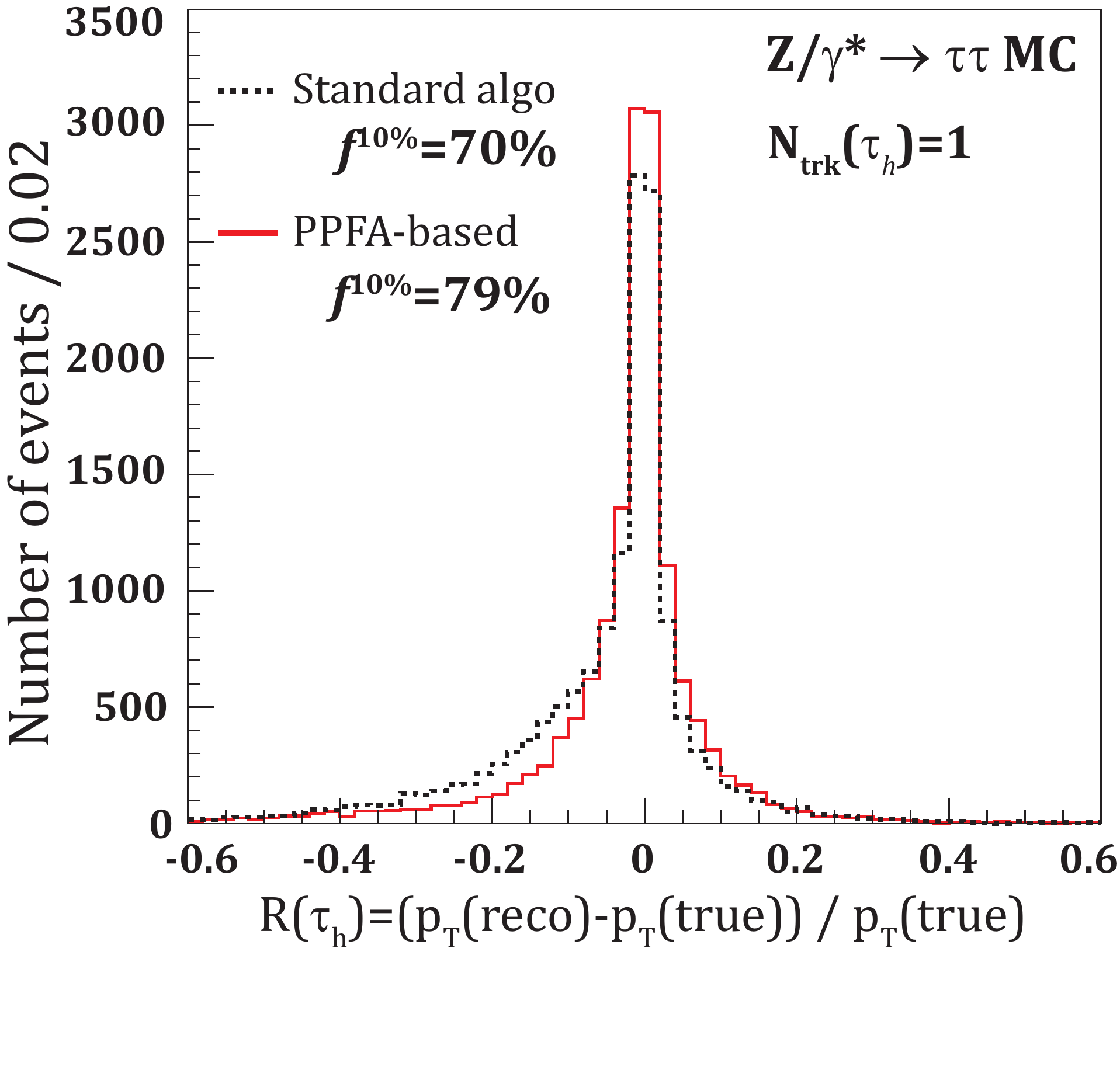}
\includegraphics[width=0.30\linewidth]{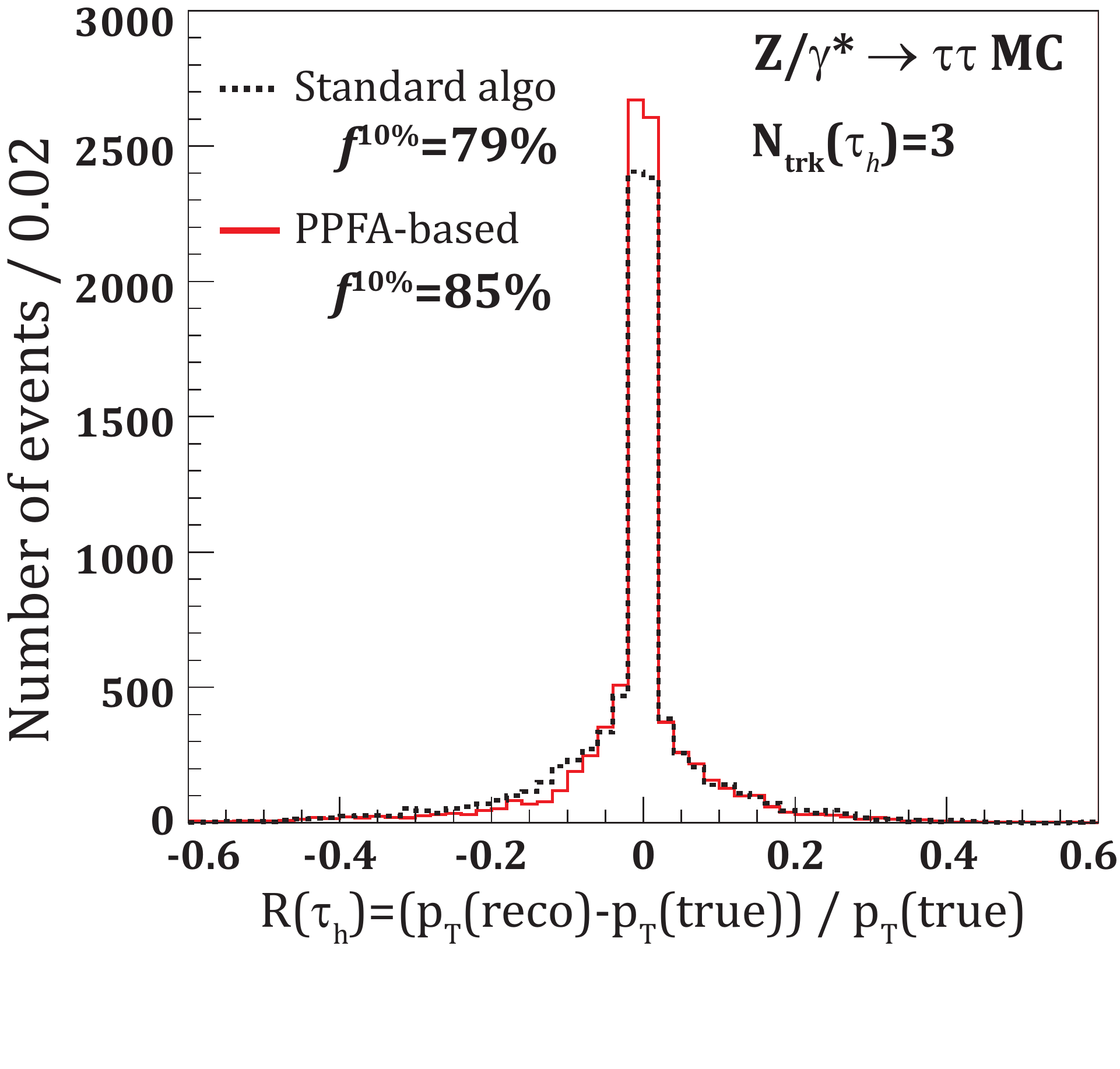}
\includegraphics[width=0.30\linewidth]{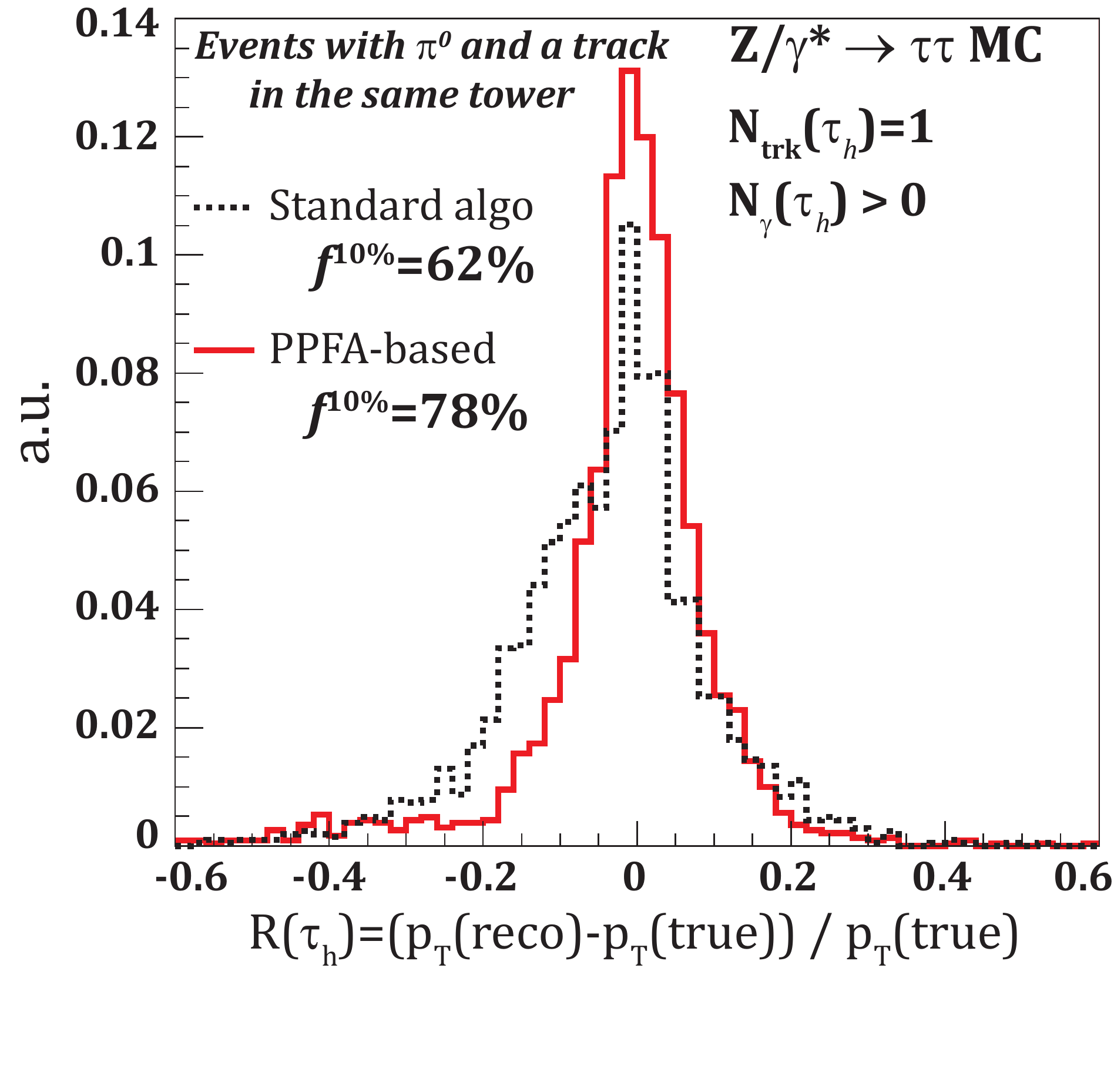}
\caption{Comparison between reconstructed transverse momentum and true transverse momentum of the hadronic tau for $\Ztt$ events in CDF II detector simulation. The red solid line corresponds to the likelihood method, the black dashed line corresponds to standard CDF tau reconstruction. (a): events with 1-prong tau; (b): events with 3-prong tau; (c): events with significant energy overlap where 1-prong tau is required to have a track and at least one reconstructed CES cluster in the same calorimeter tower.}
\label{TauReco}
\end{figure}

While our procedure has been tuned to improve the overall energy resolution and not necessarily to reconstruct and identify each individual particle in the most optimal fashion, we analyze the same sample of simulated $Z \to \tau \tau$ events with minimal reconstruction requirements to assess whether the $K^0_L$ correction implemented in the algorithm performs as expected. We find that among the hadronically decaying tau candidates, for which the $K^0_L$ correction has been invoked by our procedure, about 80\% of the candidates indeed contain a genuine $K^0_L$ in the tau decay chain based on the generator level information. In the remaining 20\% of the cases, the application of the correction has been triggered by either significant fluctuations in the hadron showering detected by the algorithm, or due to various rare mistakes in the baseline tau candidate reconstruction. Some of these mistakes, e.g. significant track momentum mismeasurements, happen very rarely, but the $K^0_L$ correction is capable of detecting at least some of these cases. In such events, the $K^0_L$ correction actually improves the overall jet energy resolution, albeit for the wrong reasons. Note that in our procedure, the $K^0_L$ correction is only applied to tau jet candidates with exactly one reconstructed charged particle track and no reconstructed photon candidates, as this configuration is the most prone to significant energy mismeasurements due to late showering energetic $K^0_L$'s. We find that of all tau candidates with a genuine $K^0_L$ falling into this category, in 60\% of the cases either a ``missing photon'' or ``missing kaon'' correction is applied in our procedure. Of these cases, about 75\% of the time the p-value reaches an acceptable value after applying the ``missing photon'' correction alone and the ``missing kaon'' correction is thus not invoked. In the remaining 25\% of the cases, the algorithm applies the ``missing kaon'' correction. The sequential application of the corrections used in our procedure is practical given that in the CDF environment the majority of mismeasurements happen due to unreconstructed photons (tagged using CES clusters). However, it is clear that one could further improve individual particle identification with the already existing PPFA tools. For example, one could make a comparison of the the p-values after applying each of the two corrections separately and then make a decision on which one is more appropriate in a particular case. Nevertheless, we conclude that the performance of the correction procedure in our algorithm is consistent with the expectation.

\subsection{PPFA Energy Resolution}
Figures~\ref{TauReco}(a) and (b) show the relative difference between the PPFA reconstructed tau jet transverse momentum and the true visible transverse momentum, obtained at generator level for one and three-prong hadronic tau jets. For comparison, the same plots show the performance of the standard CDF tau reconstruction  (see~\cite{ZttPaper} for details) shown as dashed line using the same simulated $Z \to \tau \tau$ events. It is evident that the PPFA algorithm has been able to converge to the correct energy without resorting to complex ad-hoc corrections used in the standard CDF reconstruction. The improvement is particularly striking 
in cases with significant energy overlaps, as illustrated in Fig.~\ref{TauReco}(c), which shows the same distribution, but for one-prong events containing at least one photon pointing to the same calorimeter tower as the track.

To quantify the level of improvement, we use the fraction of jets with the reconstructed energy falling within 10\% of the true jet energy, denoted as $f^{10\%}$ in Fig.~\ref{TauReco}. On average, the PPFA increases $f^{10\%}$ by about 10\%. The PPFA jet energy resolution distribution also has a more symmetric shape around the true energy and a reduced tail due to jets with
underestimated reconstructed energy. The improvement in the tail behavior is more pronounced for one-prong jets as one-prong taus more frequently contain neutral pions with significant contribution towards the total visible jet energy.

It would have been interesting to quantify the improvement in the confusion term in the energy resolution. However, disentangling the contributions to energy resolution from the confusion in assigning energy is not straightforward in PPFA due to the complex convolution of multiple detector responses, including not only the calorimeters but also the Shower Maximum detector. A qualitative feel for the level of the improvement can be deduced from the distributions shown in Fig.~\ref{TauReco}. As tau decay products rarely contain $K^0_L$'s, the shape of the underlying broad distribution beneath the the near-gaussian narrow core has a substantial contribution from the ``confusion'' cases where the electromagnetic deposition of charged pions was not correctly estimated, leading to relatively large mismeasurements. It also has other non-negligible contributions, for example large non-uniformity effects in the response of the electromagnetic calorimeter near the edges of the towers can lead to substantial mismeasurements in tau jet energy if an energetic photon enters the calorimeter near the edge of the tower. These additional effects do not allow one to unambiguously associate the entire shape of the broad component of these distributions with the confusion term. However, narrowing of this distribution in the PPFA case compared to the standard CDF PF-based tau jet reconstruction is most certainly due to improved treatment of the ``confusion prone'' cases, e.g. see Fig.~\ref{TauReco}(c) where the fraction of these potentially difficult events is enhanced by the requirement of a track and a CES cluster in the same tower. Therefore the observed improvement can arguably be taken as a lower bound on the relative improvement in the confusion term in PPFA as compared to the standard CDF technique. 
\section{PPFA Performance Tests Using CDF Data}

While the simulation studies show that the PPFA provides an accurate measurement in a single, self-consistent framework free of complex ad-hoc corrections, it is important to validate the algorithm performance in a realistic analysis setting using actual data. Energy resolution for hadronic tau jets cannot be evaluated directly using data. Unlike the case of $Z \to ee$ or $Z \to \mu \mu$ events where lepton momentum resolution can be inferred from the broadness of the dilepton mass spectrum, there is no such ``standard candle'' for taus  at hadron colliders. In the case of $Z \to \tau \tau$, which is the only fairly clean physics signal enriched with true taus accessible at hadron colliders, the shape of the invariant mass distribution calculated using visible tau decay products is very broad, as partial cancellation of the missing transverse energy $\met$ associated with momenta of the neutrinos from tau decays precludes reconstructing neutrino momenta. In addition to the improved energy resolution, the PPFA can potentially deliver other advantages, e.g. a better discrimination against QCD multi-jet backgrounds due to ``sharper'' shapes of identification variables and the new PPFA specific handles, such as the estimate of the jet energy uncertainty on a jet-by-jet basis and the $p$-value. However, as many of these potential improvements are correlated, disentangling and quantifying each of these potential improvements separately is not practical.  Incidentally, a sample of hadronic taus with purity suitable for such studies would have insufficient statistics due to very harsh cuts required to reduce background contamination. 

Given the above limitations, we validate the PPFA in a realistic data setting and evaluate its performance as follows. First, we demonstrate that the PFFA-based tau jet energy measurement in the data is well described by the simulation. Similarly, we show that the PPFA $p$-value is well reproduced in the data. Second, we study the tau jet invariant mass distribution for events with tau decays dominated by $\tau \to \rho \nu \to \pi^+ \pi^0 \nu$ and compare the PPFA-based measurement with that obtained using standard CDF reconstruction. While such invariant mass is only moderately sensitive to the jet energy resolution, this test allows an indirect validation of the PPFA jet energy resolution and a comparison with the standard CDF reconstruction. Finally, as a qualitative demonstration of the PPFA potential for enhanced background discrimination, we perform two side-by-side proto-analyses using similar data selections that rely on discriminators provided by the PPFA in one case and the standard CDF reconstruction in the other. 

\begin{figure*}[htb]
\includegraphics[width=0.48\linewidth]{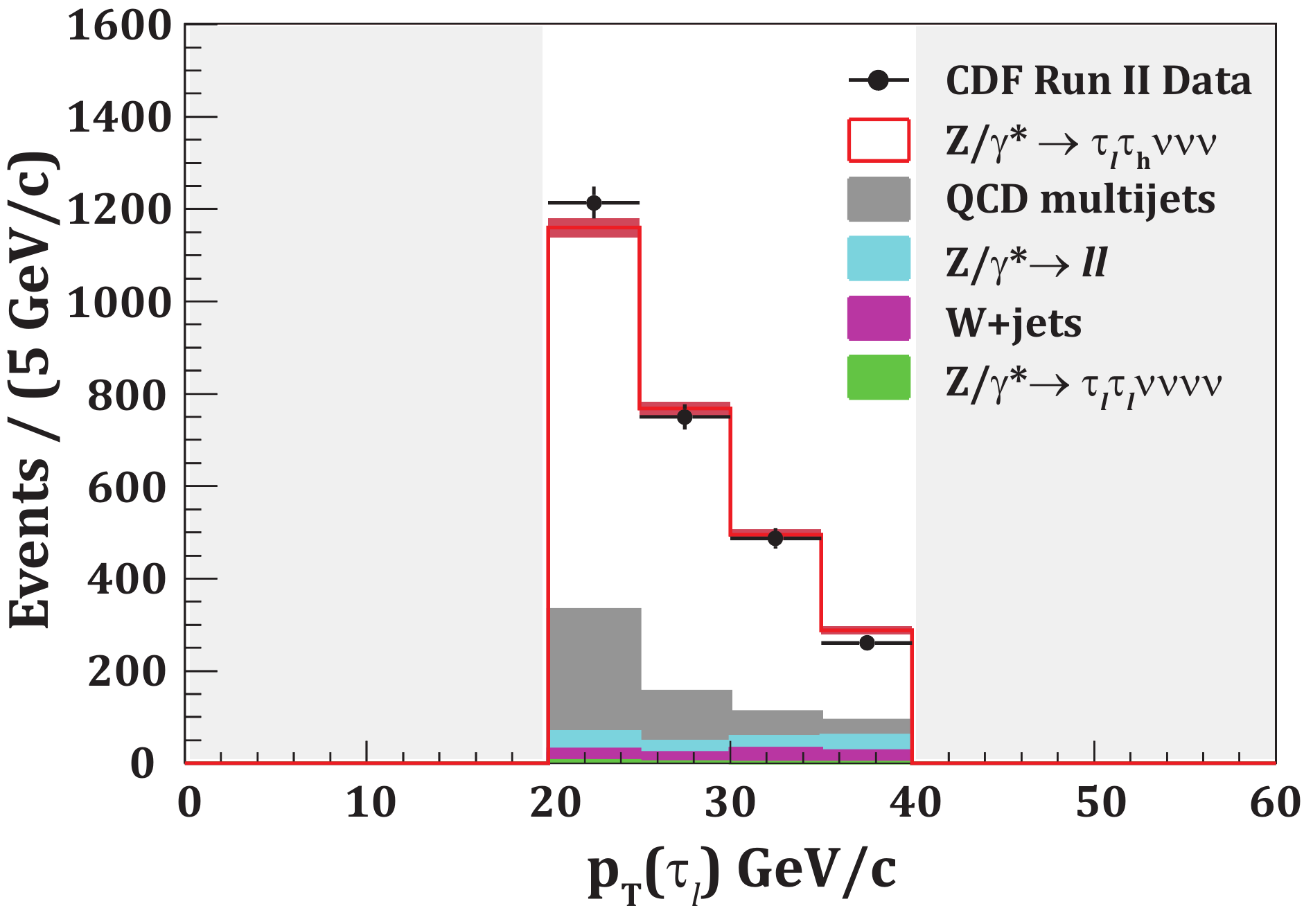}
\includegraphics[width=0.48\linewidth]{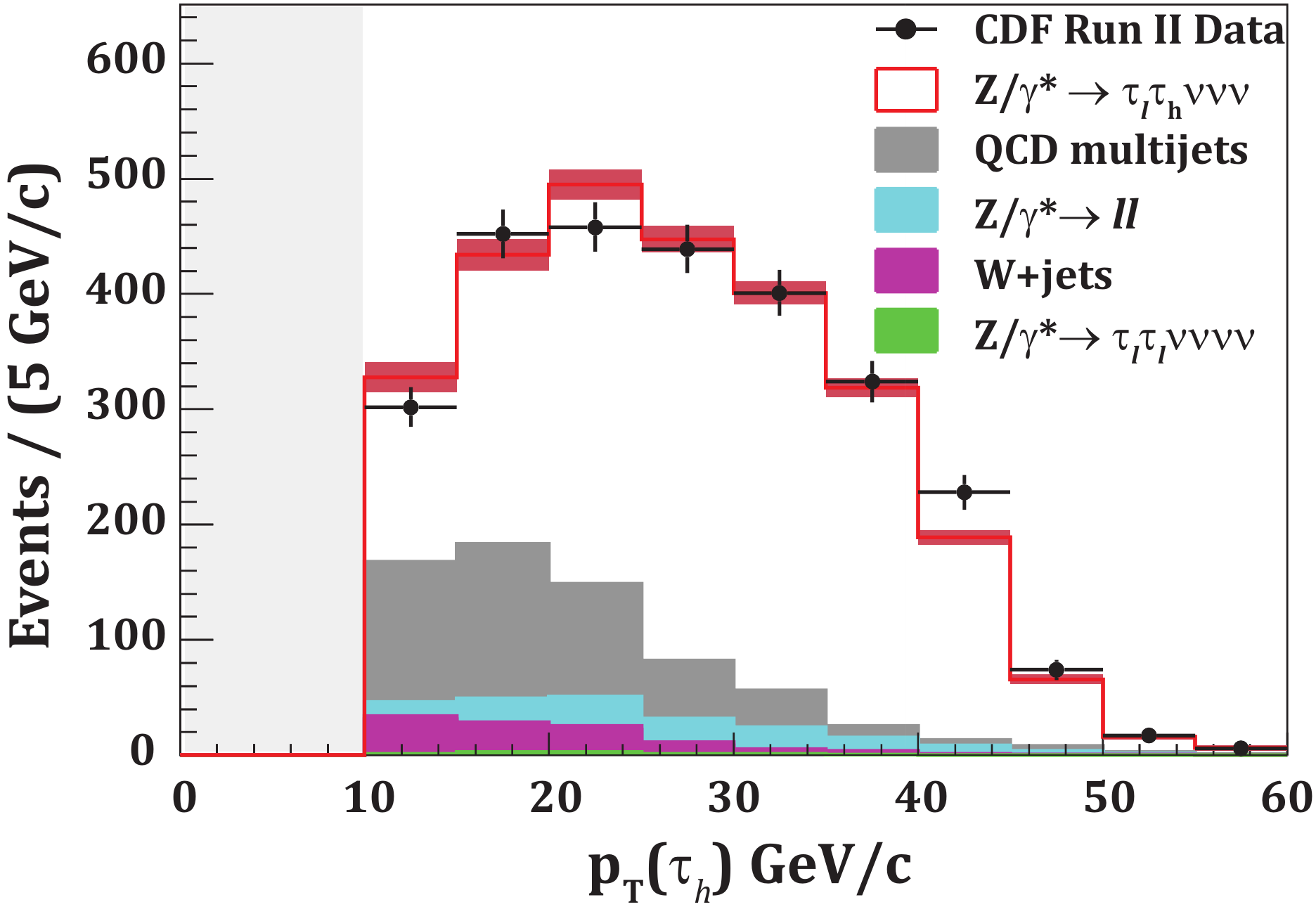}
\caption{Kinematic distributions demonstrating the purity of the clean tau sample after $Z \to \tau \tau \to l \tau_h \nu \nu \bar{\nu}$ ($l=e$ or $\mu$) events are extracted from CDF data with tight selection requirements: (a) transverse momentum of the light lepton, (b) transverse momentum of visible decay products of the hadronically decaying tau lepton, $\tau_h$.
\label{fig:data_kinematics_clean}}
\end{figure*}

\subsection{Validation of the PPFA Reconstruction Using $Z \to \tau \tau$ Data}

We use a fairly clean and well understood sample of $Z \to \tau \tau$ events collected by CDF in Run-II in the channel where one tau lepton decays hadronically ($\Tthnu$) and the other decays to a light lepton ($\tau \to l \nu_\tau {\bar\nu}_l$ where $l$ stands for an electron or muon). First, we require a tightly isolated reconstructed muon or an electron with $20<p_T<40$ GeV/c and a hadronic tau jet candidate selected with loose identification requirements. Second, tau jet candidates are required to have a seed track with $p_T>10$  GeV/c; no explicit requirement on the full momentum of the jet is applied to exclude biases owing to the choice of a tau energy reconstruction algorithm. Following this, several event topology cuts are applied to reduce contamination due to cosmics, $Z/\gamma^* \to ee$, $Z/\gamma^* \to \mu\mu$ and $W+$jets events. A full list of selections is available in~\cite{mmc}. The remaining QCD multi-jet background is estimated from data, using events with lepton and tau candidates having electric charge of the same sign. We rely on simulation to estimate $Z/\gamma^{*} \to  \tau \tau$ , $\Zee$, $\Zmumu$ and $W+$jets contributions. These processes are generated using Pythia Tune A with CTEQ5L parton distribution functions~\cite{cteq} and the detector response is simulated using the GEANT-3 package~\cite{geant3}. 

\begin{figure*}[t]
\includegraphics[width=0.48\linewidth]{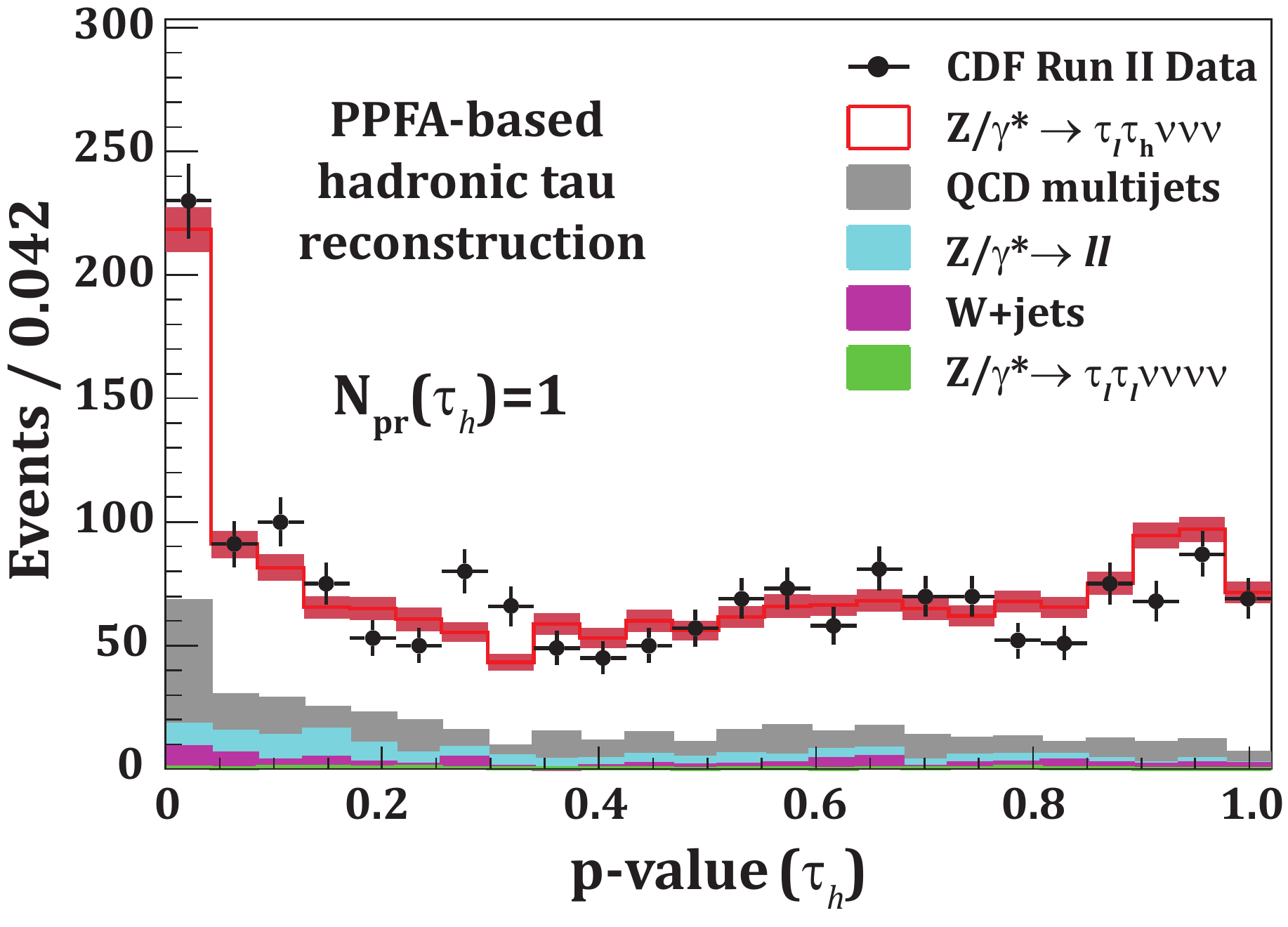}
\includegraphics[width=0.48\linewidth]{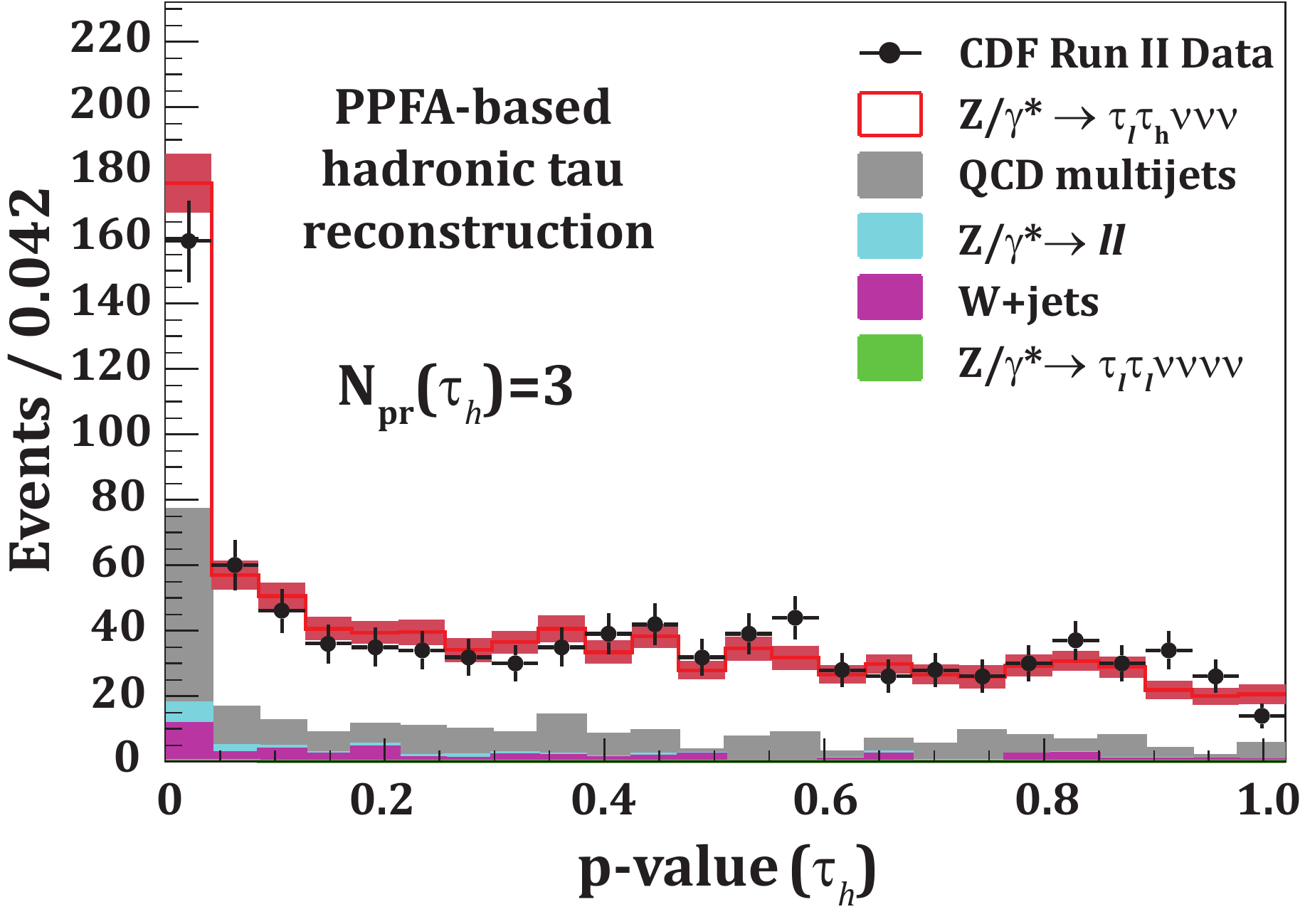}
\caption{Distribution of hadronic tau candidate p-value for events passing selection requirements in data (points) compared to the sum of background and signal predictions. Left : 1-prong taus. Right: 3-prong taus.
\label{fig:data_pvalue}}
\end{figure*}

Once the sample is selected, the PPFA reconstruction is performed in data and simulation. A thorough comparison of kinematic distributions sensitive to the hadronic tau jet energy measurement has allowed us to conclude that the PFFA performance in the data is well described by the simulation. As an illustration, Figs.~\ref{fig:data_kinematics_clean}(a) and (b) show lepton momentum and PPFA-based hadronic tau jet momentum distributions for the selected $Z/\gamma^{*}\to\tau\tau$ candidate events to demonstrate the good agreement between data and simulation, as well as to give readers a feel of the purity of the sample used.

As for the new handles made available by the PPFA, we particularlly studied the reduced $p$-value, which quantifies the level of consistency of the contributing calorimetric measurements with the hypothesis maximizing the PPFA likelihood. Despite its seeming complexity, the distribution for the reduced $p$-value is well described by the simulation. Figure~\ref{fig:data_pvalue} shows the distribution of the PPFA $p$-value for selected hadronic tau candidates with one or three charged tracks. Apart from the good agreement between the data and simulation, it is evident that the reduced $p$-value provides discrimination against the jets from multi-jet QCD events and can be utilized in physics analyses to improve the purity of selected data.

\subsection{PPFA Energy Resolution}

As discussed earlier, a direct measurement of the energy resolution for hadronic tau jets using data is not possible as the presence of multiple neutrinos in the event precludes reconstruction of the $Z$ boson mass. Conventional estimators performing partial reconstruction of the mass, e.g. the transverse mass of the lepton, hadronic tau jet and the missing transverse energy, all result in broad shapes owing to the unreconstructed neutrinos. The width of these distributions is nearly independent of the tau jet energy resolution\footnote{Even in more advanced approaches designed to improve mass reconstruction for ditau resonances, e.g. the MMC technique~\cite{mmc}, the resolution is still dominated by the accuracy of the missing transverse energy measurement.}, precluding quantitatve estimations of the latter from data.

\begin{figure*}[t]
\includegraphics[width=0.48\linewidth]{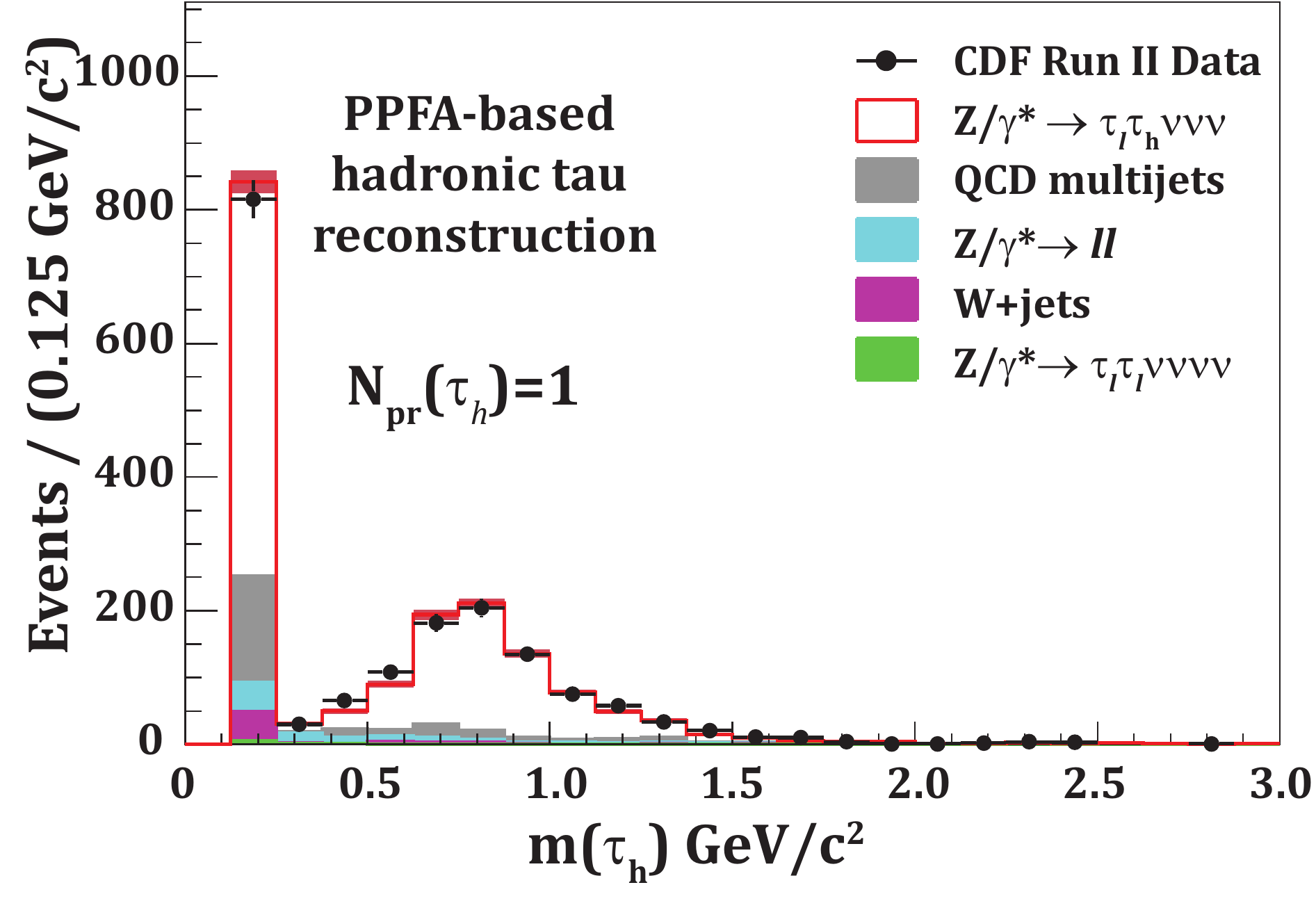}
\includegraphics[width=0.47\linewidth]{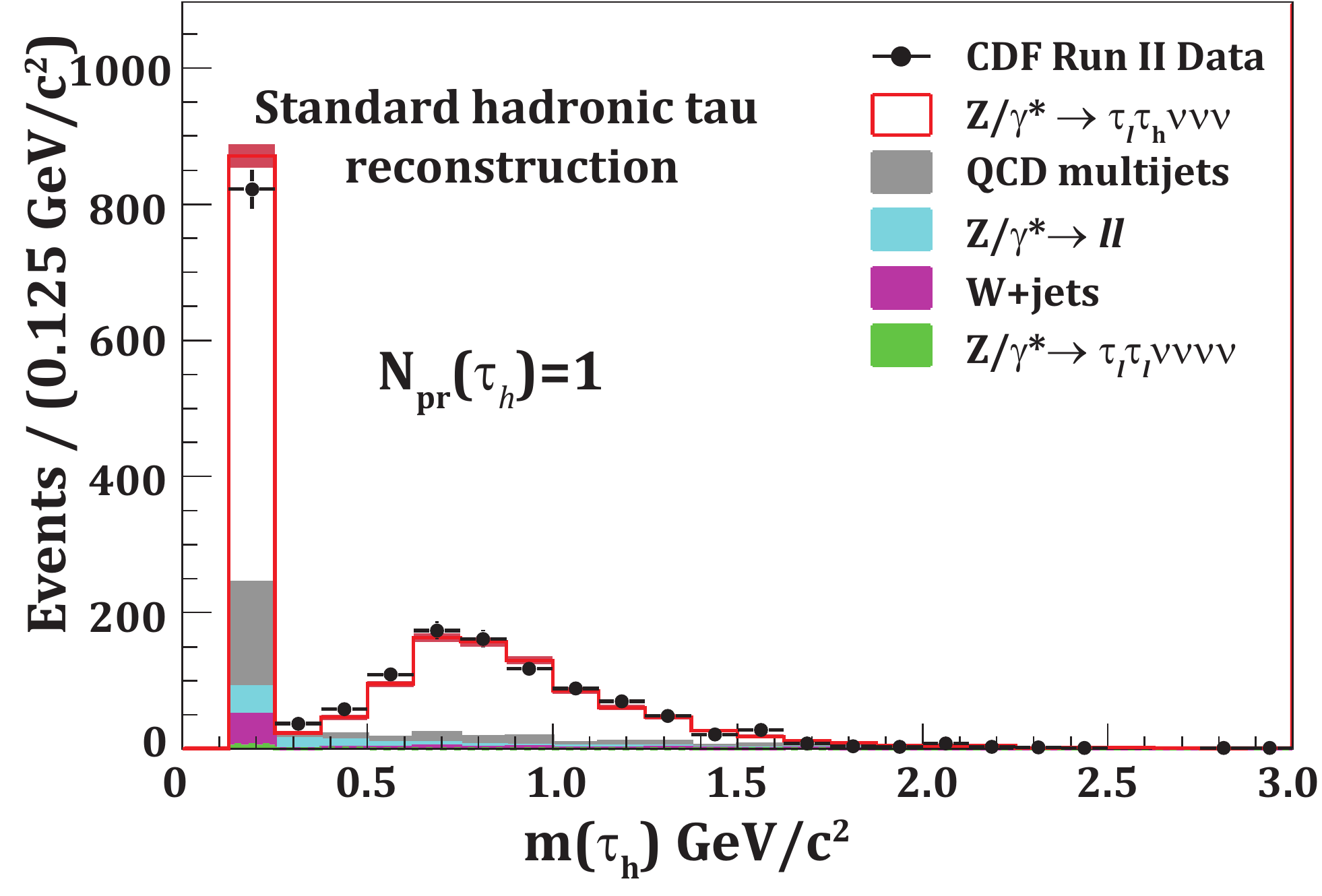}
\caption{Distribution of the invariant mass for reconstructed hadronic tau candidates in the clean $Z \to \tau \tau \to l \tau_h \nu \nu \bar{\nu}$ ($l=e$ or $\mu$) sample extracted from CDF data using tight selection requirements: comparison between PPFA reconstruction (a) and standard CDF reconstruction (b) for all 1-prong tau candidates. 
\label{fig:data_mass}}
\end{figure*}

Although only modestly sensitive to the accuracy of the jet energy measurement, the reconstructed invariant mass of the constituents of a tau jet can be used for qualitative comparisons. In particular, a significant fraction of one-prong tau jets is produced in decays $\tau_h^{\pm} \rightarrow \nu_{\tau} \rho^{\pm}(770) \rightarrow \nu_{\tau} \pi^{\pm} \pi^0$. In these decays, the invariant mass of the hadronic tau jet should be consistent with the mass of $\rho$-meson and the width of the distribution is sensitive (although somewhat weakly) to the resolution of the hadronic tau jet energy measurement. Figures~\ref{fig:data_mass}(a) and (b) show distributions of the invariant mass of the one prong tau candidates reconstructed in the data using the PPFA approach and the standard CDF tau reconstruction, with the simulation predictions overlaid. Note that the pedestal near $m=0.14$ GeV/$c^2$ is due to tau jets with no reconstructed photons, which includes $\pi^0$-less one-prong tau decays as well as the cases with the photon being unreconstructed. While these comparisons do not allow quantifying the improvement in the jet energy measurement resolution, it is evident that the PPFA technique provides a better measurement of the tau invariant mass. Similar improvements can be expected for other measurable quantities related to particle and energy flow within a tau jet candidate. As tau identification mainly relies on exploring differences in particle and energy flow properties between narrow tau jets and the broader generic jets from the QCD multi-jet backgrounds, such improvements have a potential of improving rejection of multi-jet backgrounds. 

\subsection{Tau Identification and Background Discrimination}

\begin{figure}[t]
\centering
\begin{tabular}{cc}
\includegraphics[width=0.48\linewidth]{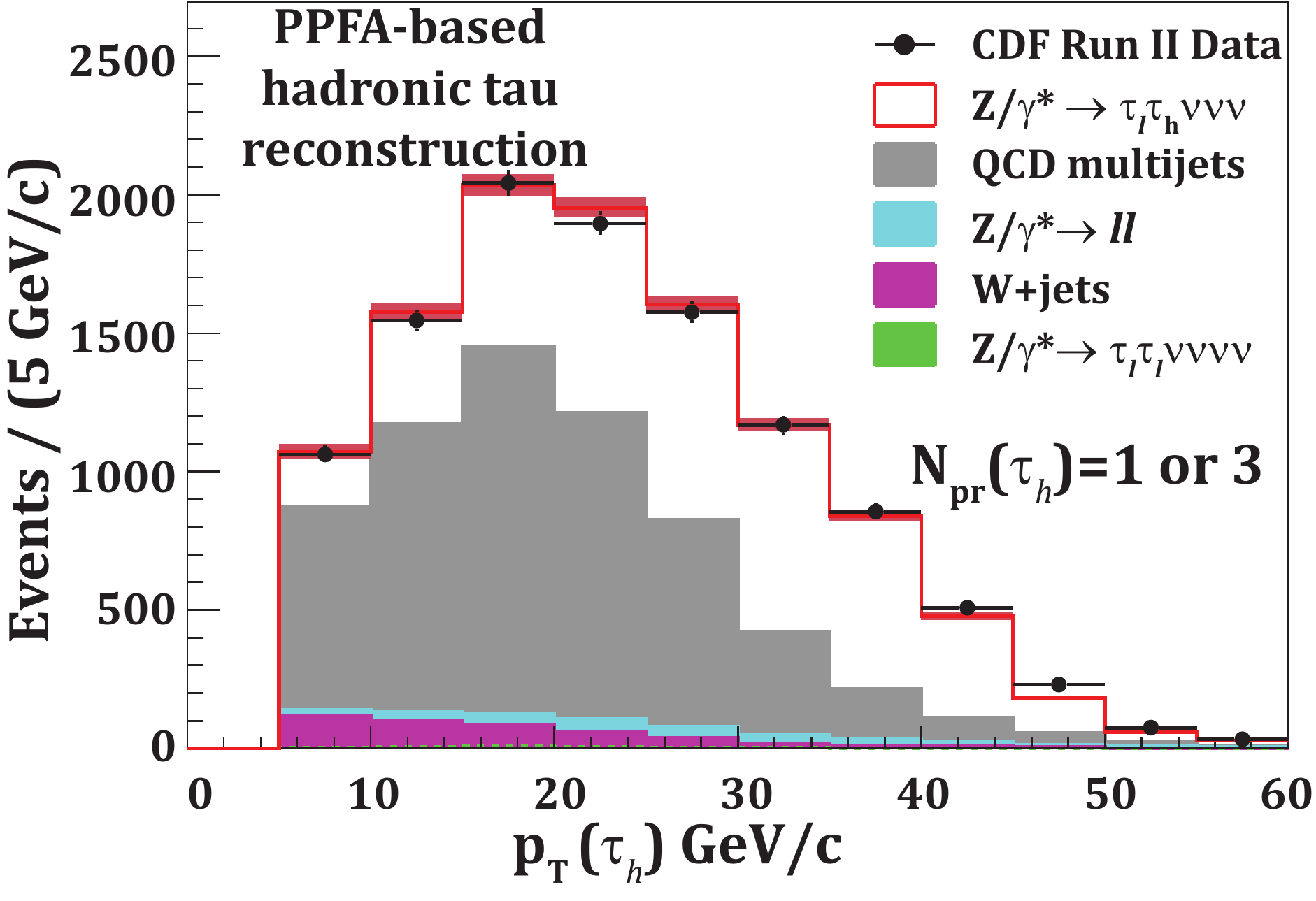} \put(-30,15){\large (a)} &
\includegraphics[width=0.48\linewidth]{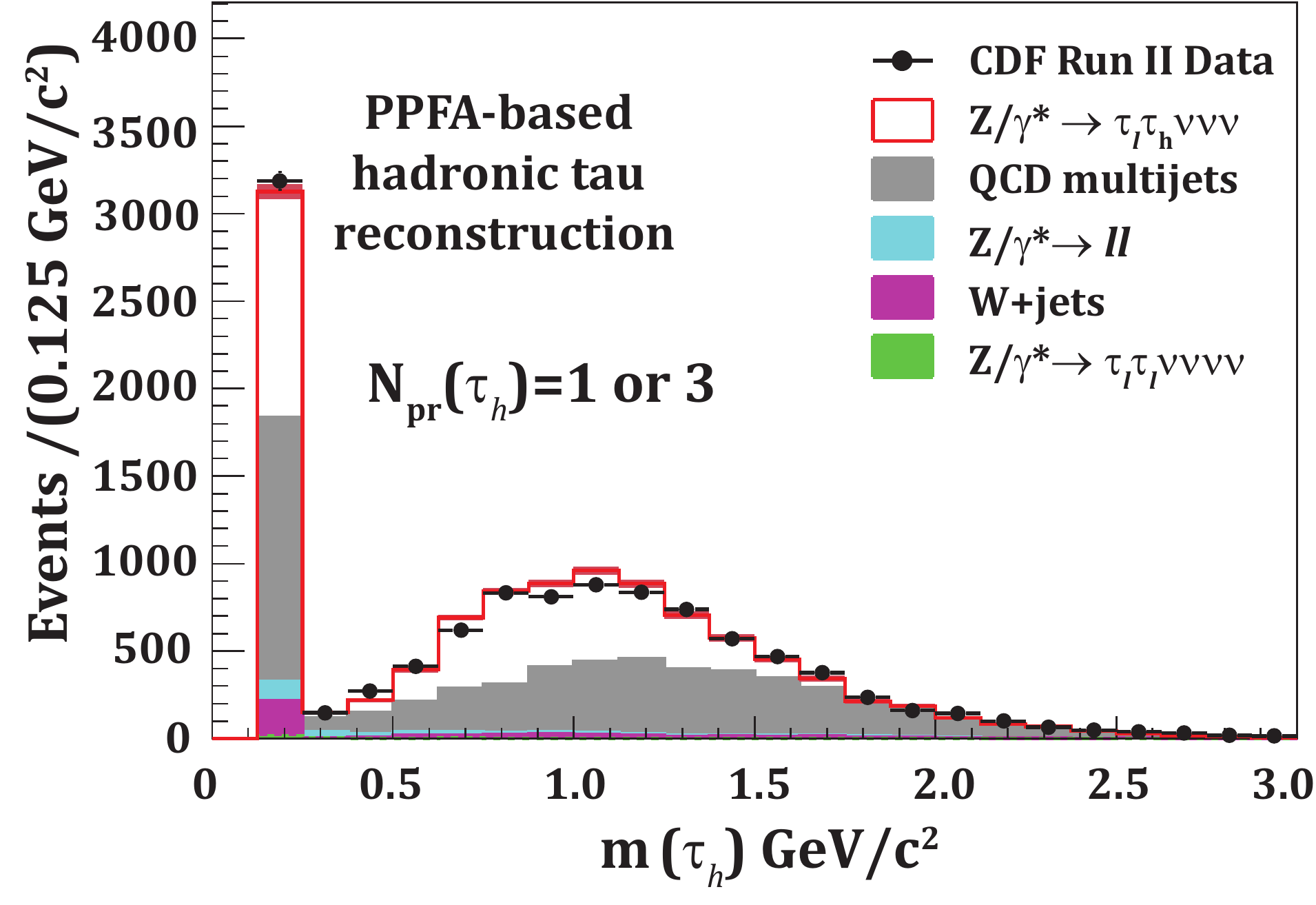} \put(-30,15){\large (b)} \\
\includegraphics[width=0.48\linewidth]{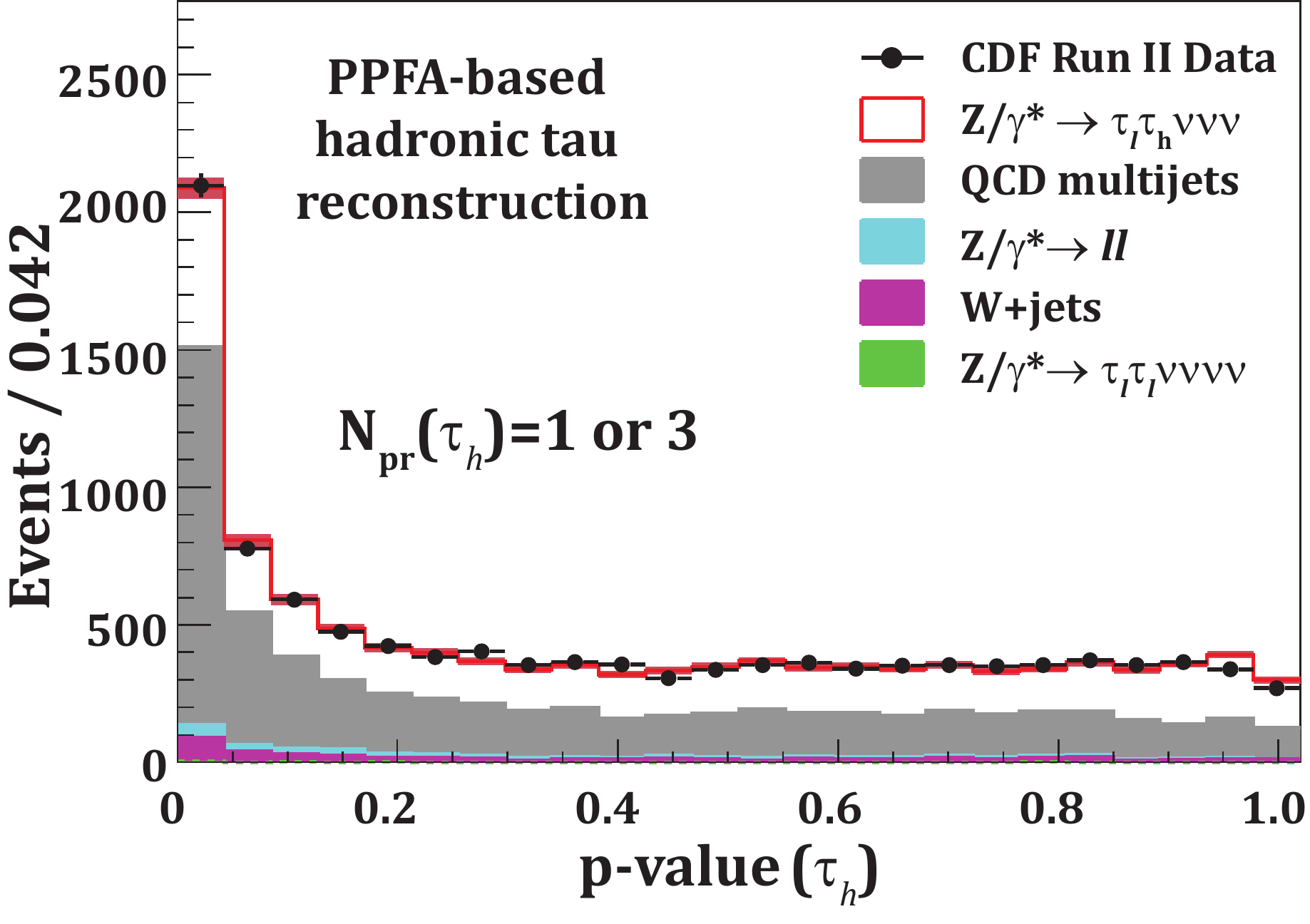} \put(-30,30){\large (c)} &
\includegraphics[width=0.48\linewidth]{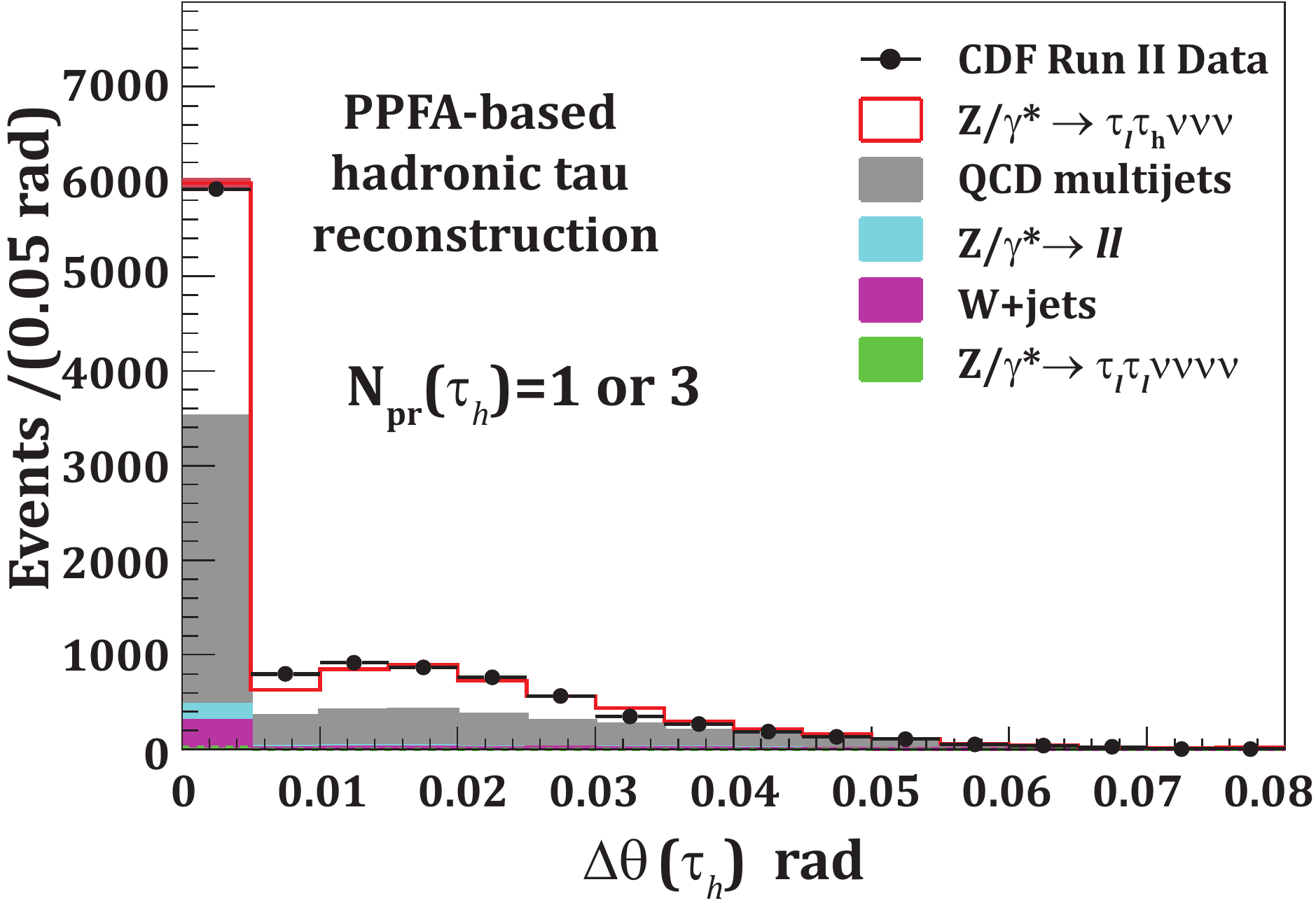} \put(-30,30){\large (d)} \\
\end{tabular}
\caption{Distributions for the selected 1- and 3-prong tau jet candidates in the sample with enhanced contribution of the QCD multi-jet events. Data (points) compared to the sum of background and signal predictions: (a) transverse momentum of visible decay products (b) hadronic tau visible invariant mass, (c) p-value distribution, (d) $\Delta \theta(\tau)$ distribution.}
\label{QCDenriched}
\end{figure}

While the primary goal of the PPFA is an accurate jet energy measurement in the high occupancy environment, it also provides additional tools that can be used in physics analyses to improve discrimination against backgrounds. Improved accuracy of the measurements of energy, particle and energy flow properties, as well as the new PPFA-specific handles, such as the $p$-value or the jet-by-jet energy measurement uncertainty, can all aid in discriminating hadronic tau jets from multi-jet QCD backgrounds. To illustrate this, we model two simple proto-analyses, both aiming to maximize the signal to background ratio for a sample of $Z \to \tau \tau$ candidate events by exploiting properties of the tau jet candidates. One of the analyses relies on variables calculated using standard CDF reconstruction and the other one relies on the PPFA calculations. Both analyses start with a sample of candidate $Z \to \tau \tau$ events with the level of background contamination due to the QCD multi-jet events that is typical for physics analyses\footnote{The clean $Z \to \tau \tau$ sample used so far features extremely tight lepton leg selections designed to achieve a high purity source of hadronic taus. While effective in reducing multi-jet backgrounds, such selections are not typical of physics analyses due to their very low signal efficiency.}. Compared to the high purity sample, the ``realistic'' sample is obtained by loosening isolation and some other tight quality requirements on the lepton leg and removing the requirement on the absence of additional energetic jets in the event. The purity and composition of this sample can be inferred from Fig.~\ref{QCDenriched} showing several kinematic and jet shape variables. 

\begin{table}[t]
\caption{Selections used in the two proto-analyses using either PPFA or standard CDF selection for hadronically decaying tau jets. The first group of selections corresponds to standard CDF selections applied first in both analyses. The second group shows additional non-standard selections using the invariant mass and the narrowness of the tau candidate's jet cluster that can be applied to both analyses. The last selection uses the PPFA p-value and is only applied to the PPFA proto-analysis. \label{proto-selections}}
\begin{center}
\begin{tabular}{cc}
\hline
1-prong & 3-prong \\ 
\hline
$p_T>10$ GeV/c & $p_T>15$ GeV/c \\ 
$p_T^{seed \; trk}>10$ GeV/c & $p_T^{seed \; trk}>10$ GeV/c \\
$N^{trk}_{iso \; cone}= 0$  & $N^{trk}_{iso \; cone}= 0$  \\
$0< m(\tau) <0.25$ or & \multirow{2}{*}{$0.8< m(\tau)< 1.4$ GeV/c$^2$} \\
$0.375<m(\tau)<1.4$ GeV/c$^2$ &  \\
$\Delta \theta<0.04$ & $\Delta \theta<0.015$ \\ \hline
\multicolumn{2}{l}{Only the PPFA-based analysis:}\\ 
$p>0.008$ if $p_T<20$ GeV/c & $p>0.06$ if $p_T<30$ GeV/c \\
\hline
\end{tabular}
\end{center}
\end{table}

Table~\ref{proto-selections} describes the selections applied. Momentum thresholds and seed track $p_T$ requirements are chosen to select a sample with an acceptable level of background while not being specific to either the PPFA or the standard CDF reconstruction. $N^{trk}_{iso \; cone}$ is the number of tracks with $p_T>1$ GeV/c in the isolation cone. $\Delta \theta (\tau)= {\sum E_i \times \theta_i / \sum E_i}$ is the weighted angular width of the jet calculated using the momenta of individual particles reconstructed in a jet,  similar to the case of the previously discussed jet invariant mass $m(\tau)$. The summation goes over particles in the jet, $E_i$ being the particle energy and $\theta_i$ is the angle between the particle and the visible 4-momentum of the tau jet. The specific cut choices for $\Delta \theta (\tau)$ and $m(\tau)$ (see Table~\ref{proto-selections}) aim at a high signal efficiency while rejecting the tails of the corresponding distributions dominated by the background events. These cuts are therefore expected to reduce background contamination, but are not optimized in any particular way. The selections discussed above can be equally applied to both the standard and the PPFA-based analyses. Finally, we apply an additional $p_T$-dependent cut on the $p$-value in the PPFA-based analysis only. The distributions for these variables using PPFA definitions are illustrated in Figs.~\ref{QCDenriched}(b), (c) and (d). 

\begin{figure}[t]
\begin{minipage}{0.34\linewidth}
\centering
\includegraphics[width=0.95\linewidth]{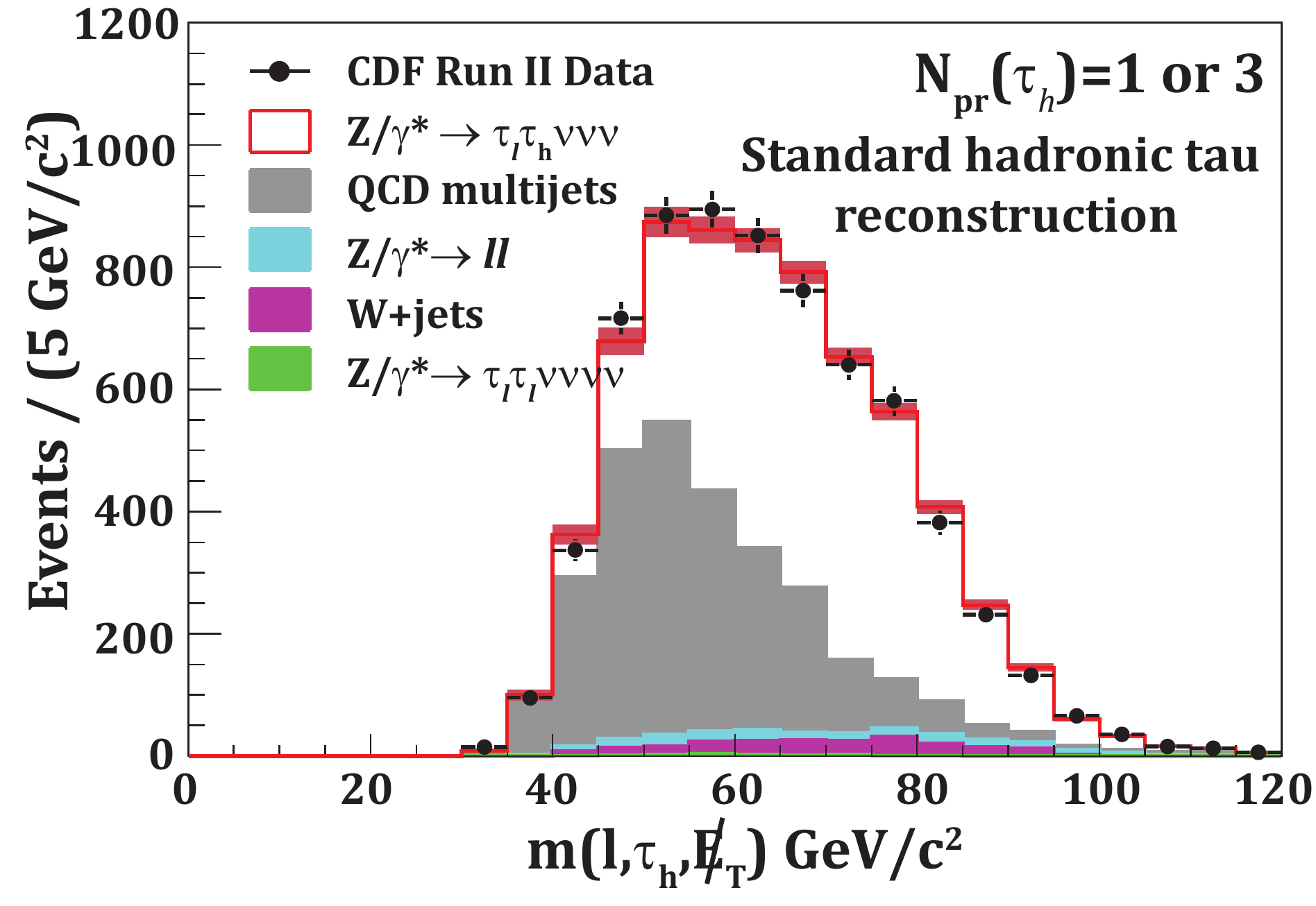}   \put(-80,22){\large (a)} \\
\includegraphics[width=0.95\linewidth]{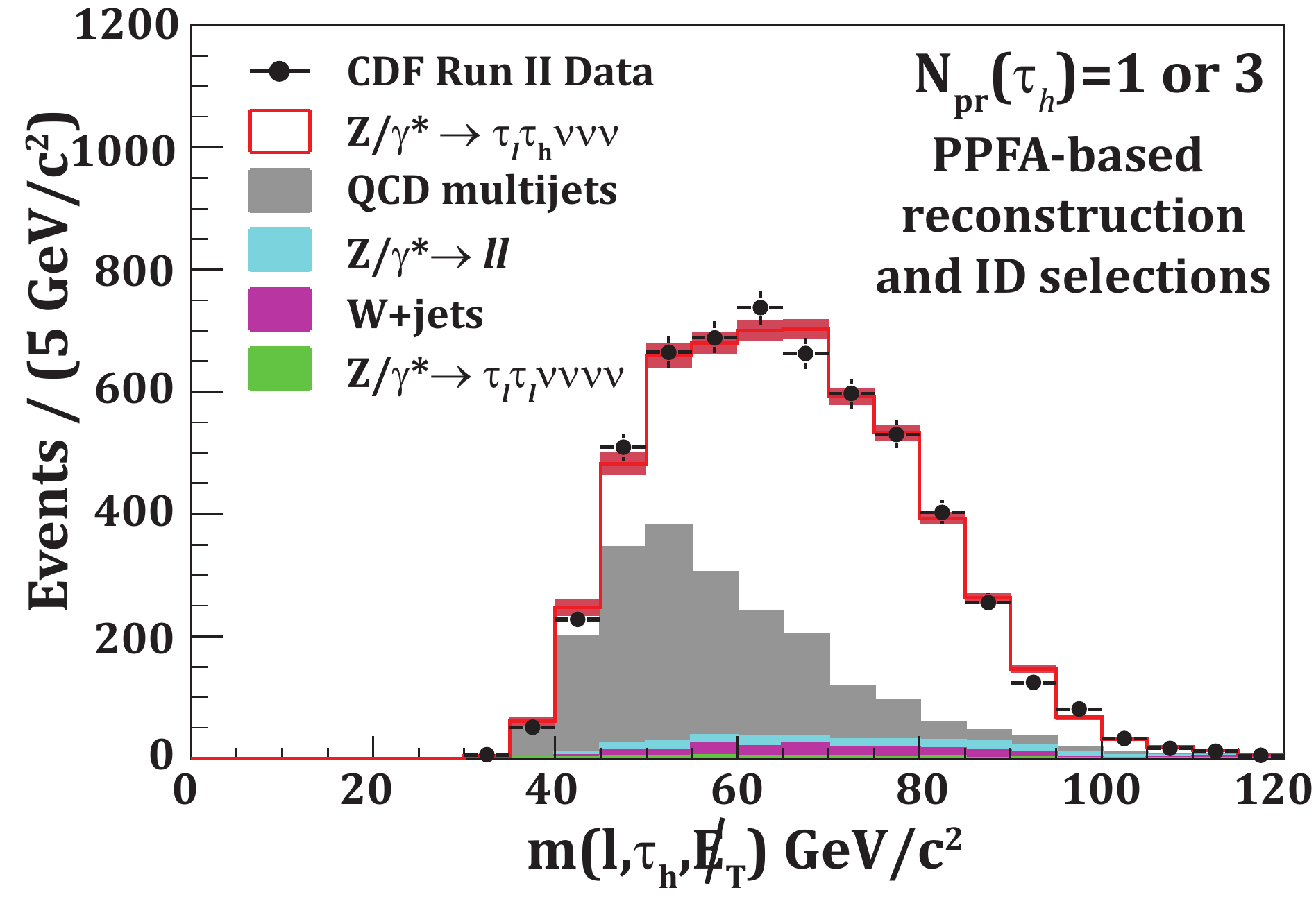}   \put(-80,25){\large (b)}
\end{minipage}
\begin{minipage}{0.66\linewidth}
\centering
\includegraphics[width=0.99\linewidth]{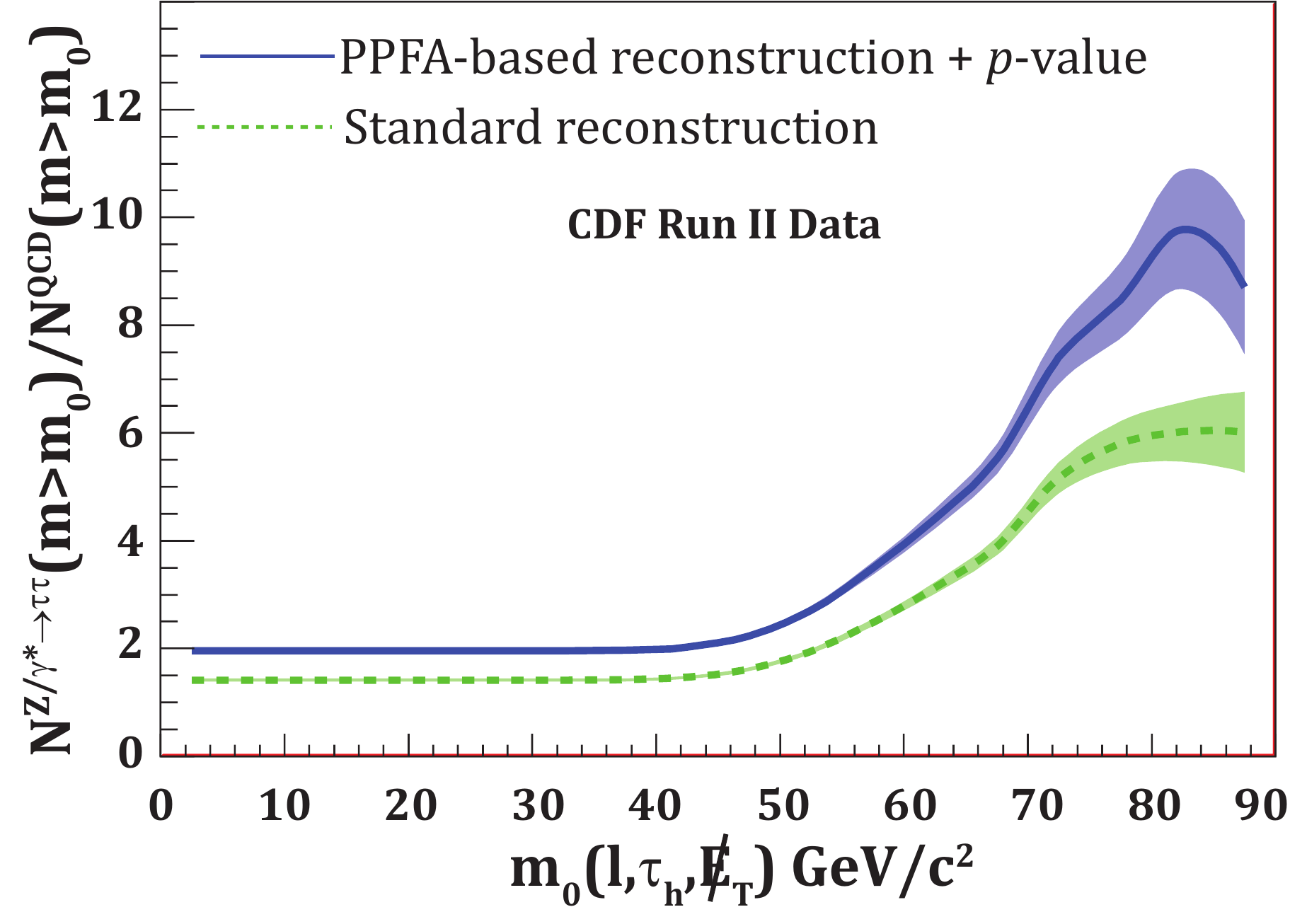}   \put(-180,80){\large (c)}
\end{minipage} 
\caption{The distribution of the visible mass, $m(\tau, l,\met)$ for events with 1 and 3-prong taus in data (points) compared to the sum of background and signal ($Z \to \tau \tau$) predictions after applying selections utilizing variables calculated using either standard CDF reconstruction (a), or PPFA (b). The PPFA case includes a cut on the p-value and otherwise the cut values are the same.  (c)  $S/B$ ratio as a function of minimal threshold $m_0$ on $m(\tau, l,\met)$. The dashed (green) line shows the standard CDF tau reconstruction and the solid (blue) line corresponds to PPFA; bands indicate the statistical uncertainty on the ratio due to the size of the sample and fluctuations in background contributions.}
\label{TauLeptMetMass}
\end{figure}

To compare the default reconstruction and PPFA side-to-side, Figs.~\ref{TauLeptMetMass}(a) and (b) show the ``after'' distributions for $m(l, \tau, \met)$, the visible mass of lepton, tau and missing transverse energy\footnote{This quantity is frequently used as the final discriminant in physics analyses~\cite{HttMSSMPaper,cdf_tau}}, for each of the two proto-analyses.  As a quantitative figure of merit for the comparison of the two techniques,  in Fig.~\ref{TauLeptMetMass}(c) we show the ratio $N^{Z \to \tau \tau} (m>m_0) /N^{QCD} (m>m_0) $, where $N^{Z \to \tau \tau} (m>m_0)$ and $N^{QCD} (m>m_0) $ are the estimated rates of events and background events with $m(l, \tau_h,\met)>m_0$, for the selected sample as a function of $m_0$. Note that backgrounds are heavily dominated by the QCD multi-jet events. Near its maximum, the $S/B$ ratio is a factor of 1.7 higher for the PPFA case. While by no means exhaustive, this comparison indicates the potential of the PPFA technique in discriminating hadronic tau jets from quark and gluon jets, thus providing a nice byproduct of the method that can be utilized in physics analyses.

\section{Conclusions}
The PPFA is a consistent, probabilistic framework designed for accurate reconstruction of the jet energy in the high occupancy environment, relevant for experiments operating in the very high luminosity regime or featuring coarse calorimeter segmentation. The framework is based on ``first principles'' and is essentially free of ad-hoc corrections. The PPFA can be implemented in a realistic detector setting, as demonstrated using the example of hadronic tau reconstruction at CDF. It is shown to provide a more accurate jet energy measurement and better discrimination against backgrounds compared to the existing tools utilizing the particle flow concept. For hadronic tau reconstruction, the new tools provided by the PPFA, such as a jet-by-jet estimate of the jet energy uncertainty and the $p$-value quantifying the likelihood of the current hypothesis about particle content of a jet, can be used to further improve energy resolution and provide better discrimination against backgrounds. The proposed technique can be utilized at the LHC experiments once the machine is upgraded for the very high luminosity regime as well as at future collider experiments.

\section*{Acknowledgments}
We thank Anthony K. Rose, Robert M. Roser, and Jeffrey K. Roe for carefully reading the manuscript and their useful suggestions. We thank our CDF collaborators for their support and the Fermi National Accelerator Laboratory, where a part of the work on the paper has been performed. This work would not be possible without the funding support of the U.S. Department of Energy, the DOE OJI program, and the State of Texas.

\def\NCA{Nuovo Cimento}
\def\NIM{Nucl. Instrum. Methods}
\def\NIMA{{Nucl. Instrum. Methods} A}
\def\NP{Nucl. Phys.}
\def\NPB{{Nucl. Phys.} B}
\def\PLB{{Phys. Lett.}  B}
\def\PRL{Phys. Rev. Lett.}
\def\RPP{Rep. Prog. Phys.}
\def\PRD{{Phys. Rev.} D}
\def\PR{Phys. Rep.}
\def\PRP{Prog. Theor. Phys.}
\def\ZPC{{Z. Phys.} C}
\def\MPL{{Mod. Phys. Lett.} A}
\def\EPJC{{Eur. Phys. J.} C}
\def\CPC{Comput. Phys. Commun.}

\end{document}